\documentclass[a4paper,12pt]{article}
\pdfoutput=1

\usepackage[english]{babel}
\usepackage{amsmath,amsfonts,amssymb,latexsym}
\usepackage{jheppub}
\usepackage{color}
\usepackage{mathbbol,mathrsfs}
\usepackage{slashed}
\usepackage{enumerate}
\usepackage[footnotesize]{caption}
\usepackage[center,footnotesize,hang]{subfigure}
\usepackage{verbatim}
\usepackage{cancel}

\newcommand{\be}{\begin{equation}}
\newcommand{\ee}{\end{equation}}
\newcommand{\beq}{\begin{equation}}
\newcommand{\eeq}{\end{equation}}
\newcommand{\bac}{\beq\begin{array}}
\newcommand{\eac}{\end{array}\eeq}
\newcommand{\ba}{\begin{array}}
\newcommand{\ea}{\end{array}}
\newcommand{\bea}{\begin{eqnarray}}
\newcommand{\eea}{\end{eqnarray}}

\newcommand{\nn}{\nonumber}
\newcommand{\raw}{\rightarrow}

\newcommand{\blue}[1]{\color{blue} #1 \color{black}}

\numberwithin{equation}{section}

\newcommand{\eps}{\epsilon}
\newcommand{\vep}{\varepsilon}
\newcommand{\la}{\lambda}

\newcommand{\LL}{\mathscr{L}}

\newcommand{\UH}{\mathbf{U}}
\newcommand{\MH}{\mathbf{M}}

\newcommand{\TL}{\mathbf{T}}
\newcommand{\VL}{\mathbf{V}}

\newcommand{\Gf}{\mathcal{G}_f}

 \def\cH{{\cal H}}

\def\cO{{\mathcal O}}

\newcommand{\MeV}{\;\text{MeV}}
\newcommand{\GeV}{\;\text{GeV}}

\newcommand{\ps}{\;\text{ps}}
\newcommand{\fb}{\;\text{fb}}

\newcommand{\re}{\text{Re}}
\newcommand{\im}{\text{Im}}

\newcommand{\hc}{\text{h.c.}}

\newcommand{\ov}[1]{\overline{#1}}
\newcommand{\tra}{\text{Tr}}
\leftline{\blue{TUM-HEP-822/11\hspace{1cm}
DFPD-11/TH/19 \hspace{1cm}FTUAM-11-69\hspace{1cm}
IFT-UAM/CSIC-11-10}}
\bigskip

\title{Minimal Flavour Violation with Strong Higgs Dynamics}
\author[a]{R.~Alonso,} 
\author[a]{M.B.~Gavela,} 
\author[b,c]{L.~Merlo,}
\author[d]{S.~Rigolin,} 
\author[a]{and J.~Yepes} 

\affiliation[a]{Departamento de F\'isica Te\'orica and Instituto de F\'isica Te\'orica, IFT-UAM/CSIC, \\
	       UAM Cantoblanco 28049, Madrid}
\affiliation[b]{Physik-Department, Technische Universit\"at M\"unchen, \\
  	       James-Franck-Strasse, D-85748 Garching, Germany}
\affiliation[c]{TUM Institute for Advanced Study, Technische Universit\"at M\"unchen, \\
  	       Lichtenbergstrasse 2a, D-85748 Garching, Germany}
\affiliation[d]{Dipartimento di Fisica ``G.~Galilei'', Universit\`a di Padova and \\
  	       INFN, Sezione di Padova, Via Marzolo~8, I-35131 Padua, Italy}

\emailAdd{rodrigo.alonso@uam.es}
\emailAdd{belen.gavela@uam.es}
\emailAdd{luca.merlo@ph.tum.de}
\emailAdd{stefano.rigolin@pd.infn.it}
\emailAdd{ju.yepes@uam.es}

\abstract{
We develop a variant of the Minimal Flavour Violation ansatz  for the case of a strongly interacting 
heavy-Higgs boson sector. The tower of effective operators differs from that for a Higgs system in the 
linear regime, and the new operators obtained at leading order include a CP-odd one. We investigate the 
impact of these operators on $\Delta F=1$ and $\Delta F=2$ observables, demonstrating that the non-linear 
scenario may have an interesting impact on the anomalies in present data.
}


\begin{document}

\maketitle

%
\section{Introduction}
%
Electroweak precision tests \cite{Alcaraz:2009jr}, when analyzed within the SM {\it and} assuming a linearly 
realized Higgs sector, clearly point since long  to a light scalar degree of freedom. By light we mean 
here and below a Higgs mass close to its present lower bound, while by heavy we will refer to a Higgs 
mass in the TeV range or larger.

Latest LHC data could be hinting to a SM-like light Higgs particle with a mass around 125 GeV~
\cite{ATLAS,CMS}. Would this hint be substantiated by LHC data in the next months, with the resulting 
picture pointing towards a pure SM scenario at those energies, the need to find alternative solutions 
to the electroweak hierarchy puzzle would be more compelling than ever: the SM ``desert" should 
flourish with new physics (NP) at the TeV scale, to solve the electroweak hierarchy puzzle.

Nevertheless, the observed excess of events in the 120-130 GeV region is still well compatible with a 
background fluctuation, in which case the two most obvious avenues will be: i) either the Higgs is 
there but hidden in the data, for instance because it mixes with singlet scalars, which is known to 
generically dilute the strength of the signals (see for example \cite{vanderBij:2006ne} and reference 
therein); ii) or the Higgs mass is around the TeV range or even larger (up to the limit of infinite mass).

Possibility ii) is a very interesting physics option, which may correspond for instance to a composite 
scalar, like in technicolor~\cite{Kaplan:1983sm,Dugan:1984hq}, or simply to the absence of a physical 
Higgs excitation. Recall that in the Standard Model of particle physics (SM), the Higgs mass is given
by  $m_H^2\sim \lambda/G_F$, where $G_F$ denotes the Fermi constant and $\lambda$ is the quartic 
self-coupling of the Higgs field. For values of the Higgs mass $m_H\ge 1$ TeV, it follows that
$\lambda\ge 1$:  the Higgs interactions enter a strong (non-linear) regime. A heavy Higgs scenario  thus 
points to non-elementary Higgs quanta and a rich strong dynamics around the TeV scale, a range 
of energy fortunately at reach at the LHC in a few years.

Notice as well that, if a light Higgs and nothing else is confirmed to exist below TeV scales, the Higgs 
sector could still encode a strong TeV dynamics to account for the hierarchy problem or, in other words, 
a non-linear realization of the electroweak symmetry breaking mechanism. For instance, recent models (see 
Ref.~\cite{Giudice:2007fh} for an overview) of a light composite Higgs sector include in the spectrum a 
light Higgs which is a quasi-Goldstone boson of the theory.

In the case of a purely weakly interacting Higgs, that is, of a purely linearly realized SM, the impact on 
low-energy observables of new physics at a scale $\Lambda\gg v$, where $v$ is the vacuum expectation 
value  (vev) of the Higgs field, can be parameterized in a model-independent way as a tower 
of effective operators organized as a Taylor series  in $v^2/\Lambda^2$. In contrast to this, in 
non-linear realizations of the Higgs mechanism the characteristic scale $\Lambda$ of the models 
is of order $v$, and a power expansion in $v^2/\Lambda^2$ is not pertinent.
A convenient model-independent way to explore then the low-energy consequences of a strong Higgs 
dynamics is to treat the massless Goldstone bosons, which are the longitudinal components of the 
$W$ and $Z$ bosons, in the same spirit that low-energy hadronic physics is described by phenomenological 
chiral Lagrangians~\cite{Nambu:1961fr,Nambu:1961tp}. In the limit in which the Higgs-boson mass goes 
to infinity, a gauged non-linear sigma model is often used to order and catalog the low-energy effects. 
The relevant symmetry protecting the longitudinal degrees of freedom of the gauge bosons is a 
global $SU(2)_L\times SU(2)_R$, spontaneously broken to the diagonal ``custodial" symmetry $SU(2)_C$ and explicitly broken by the $U(1)_Y$ gauge interaction and by the (different) masses of fermions in each $SU(2)_L$ fermion doublet.  The longitudinal degrees of freedom of the gauge bosons are 
more efficiently described in a compact way by a unitary matrix $\UH$ transforming as a $(2,2)$ 
of the global symmetry group with $\UH^\dagger \UH = \UH \UH^\dagger=\mathbb{1}$. 

The tower of effective operators that parameterizes generically the low-energy impact of a non-linear Higgs 
sector is well known to differ~\cite{Appelquist:1980vg,Longhitano:1980tm,Longhitano:1980iz} from that 
of a purely  linear regime. Those operators containing a Higgs field $\Phi$ in the purely linear regime, 
and weighted in the Taylor expansion as powers of $\Phi/\Lambda$, may not be present. Instead, the 
Goldstone boson contribution encoded in $\UH$ does not exhibit such a mass or scale suppression, as $\UH$ 
is dimensionless: much as the emission of multiple gluons is not suppressed in the non-perturbative regime 
of QCD because the interaction strength is of $\cal O$(1), the emission of the fields encoded in $\UH$ is 
not suppressed either. As a result the ``chiral" dimension of the effective operators describing the 
dynamics of the scalar sector differs from that in the purely linear regime. Typically, an operator 
containing the Higgs field in the linear expansion has its non-linear counterpart  as a lower dimension 
operator, while the operators made out exclusively of gauge and/or fermion fields keep the same dimension. 
The consequence is that the leading terms of the two expansions may and will differ.

Assuming  a  strongly interacting heavy-Higgs-boson sector and armed with the tools mentioned above, we 
develop in this work a variant of the {\it Minimal Flavour Violation} (MFV)~\cite{D'Ambrosio:2002ex} ansatz. 
The latter is not a model of flavour, but rather a general framework, where the flavour violation in the SM and beyond is described in terms of the known fermion mass hierarchies and mixings. A relevant outcome is that, while in a completely general beyond SM set-up the scale of new physics should be heavier than hundred TeV \cite{Isidori:2010kg}, in the MFV framework this energy scale is lowered to few TeV. 

The MFV ansatz builds upon the fact that, in the limit of zero fermion masses, the SM exhibits a global 
flavour symmetry 
\beq
\mathcal{G}_f =SU(3)_{Q_L}\times SU(3)_{U_R}\times SU(3)_{D_R}\,
\label{3famsym}
\eeq
plus three extra $U(1)$ factors corresponding to  baryon number, hypercharge and the Peccei-Quinn symmetry 
\cite{Peccei:1977hh,Peccei:1977ur}.  The non-abelian subgroup $\Gf$  controls the flavour structure of the 
Yukawa matrices, and we focus on it for the remainder of this paper. Under $\Gf$, the $SU(2)_L$ quark 
doublet, $Q_L$, and the $SU(2)_L$ quark singlets, $U_R$ and $D_R$, transform as: 
\beq
Q_L\sim(3,1,1)\,,\qquad\qquad
U_R\sim(1,3,1)\,,\qquad\qquad
D_R\sim(1,1,3)\;.
\label{QuarksTrans}
\eeq
The MFV hypothesis  is a  highly predictive working frame in which: i) it is assumed that, at low energies, 
the Yukawa couplings are the only sources of flavour and CP violation both in the SM and in whatever may be 
the flavour theory beyond it, which is consistent with the non-observation of flavour effects other than 
those predicted in the SM; ii) it is exploited the flavour symmetry which the SM exhibits in the limit of 
vanishing Yukawa couplings, and which is implicitly assumed to be respected by the BSM theory of flavour.

The SM Yukawa interactions, 
\beq
\LL_Y=\ov{Q}_LY_DD_RH+\ov{Q}_LY_UU_R\tilde{H}+\hc \,,
\label{YukawasL}
\eeq
break explicitly both the custodial symmetry $SU(2)_C$ and the flavour symmetry group $\Gf$~\footnote{In 
the present proposal we do not promote the custodial symmetry to be a true symmetry of the theory. This is 
particularly relevant, because imposing it the flavour symmetry of the kinetic terms would be $SU(3)^2$ 
and not $SU(3)^3$, as we are considering.}. The technical realization of the MFV ansatz promotes the up 
and down Yukawa couplings $Y_{U,D}$ to be spurion fields \cite{Feldmann:2009dc,Alonso:2011yg} which 
transform under $\Gf$ as
\beq
Y_U\sim(3,\ov{3},1)\;,\qquad\qquad 
Y_D\sim(3,1,\ov{3})\;,
\label{spurions}
\eeq
recovering the invariance under $\Gf$ of the full SM Lagrangian and also under $SU(2)_L\times SU(2)_R$ except for the 
$U(1)$ gauge interactions. Using these transformation properties, it is possible to write an effective 
Lagrangian invariant under $\Gf$, in which the flavour transformation of each operator is compensated 
by that of its operator coefficient, which is constructed with this aim out of the Yukawa spurions. The description given so far applies both to the case of a weakly interacting Higgs field, as in Ref.~\cite{D'Ambrosio:2002ex}, and to the case where a strong interacting sector is present at the TteV scale. At our knowledge, there is only one example \cite{Redi:2011zi} of strong dynamics theory at the 1 TeV scale that leads to exact MFV, even if under certain assumptions. We are not attempting, here, to work on this new physics scenario, but rather to study the results of applying the MFV ansatz on this context. In both the weakly and strongly interacting cases, the only requirement is that the low-energy effects undergo the MFV ansatz, as stated above. Within this framework, we determine below the tower of flavour-changing operators for a variant of the MFV ansatz corresponding to the case of a strong interacting Higgs regime. 
One of the novelties to be discussed is the appearance 
at leading order of a CP-odd operator, whose coefficient is necessarily complex. Notice that in the 
``minimal" version of MFV the Yukawa couplings are assumed to be the only source of the breaking 
of both $\Gf$ {\it and} the CP symmetry, but this does not need to be the case, as pointed out in 
Refs.~\cite{Colangelo:2008qp, Mercolli:2009ns, Kagan:2009bn, Ellis:2010xm}, and flavor-diagonal CP-odd phases can be 
allowed in generic MFV models. This is precisely what happens naturally at leading order in the strongly 
interacting Higgs scenario considered here, and the operator found may have a strong impact on CP-odd 
observables for the $B$ system. 

More in general, in addition to write the full MFV effective Lagrangian at leading order for the case 
of a strongly interacting Higgs, we explore  the phenomenological limits and consequences of the MFV 
tower of operators that we derive, and compare with the analysis in the literature for the case of a 
Higgs in the linear regime. At this stage, our phenomenological analysis will concentrate on the 
following observables: 
\begin{itemize}
\item[-] 
$\Delta F=1$ processes, which are sensitive to modifications of the fermion-$Z$ couplings. 
$B\raw \tau^+ \nu$ and the rare heavy meson decays  $B\raw \mu^+ \mu^-$, $B\raw X_s \,\ell^+ \,\ell^-$, 
$K^+\raw \pi^+ \,\bar\nu \,\nu$ will be most relevant to the present analysis;
\item[-] 
$\Delta F=2 $ processes, which are very sensitive  to modifications of the fermion-$W$ couplings. 
Meson oscillations and semileptonic asymmetries will be discussed, and in particular the quantities 
$\Delta M_{B_{d,s}}$, the CP-asymmetry $S_{\Psi K_S}$ in  the decay $B_d\raw \Psi K_S$, 
the CP-asymmetry $S_{\Psi \phi}$ in the decay $B_s\raw \Psi \phi$, $\epsilon_K$ and the like-sign 
dimuon charge asymmetry of semileptonic $B$ decays.
\end{itemize}
The operators considered below will produce additional flavour-conserving effects. In the spirit of MFV, 
we will not include such a discussion here, deferring the extended analysis to a future work 
\cite{USworkinprogress}.

It is worth mentioning that the results presented here, besides applying to heavy Higgs scenarios, 
could {\it a priori} be valid also in theories with both  strong interacting Higgs dynamics and a light Higgs. 
For instance, in composite Higgs models all degrees of freedom of the Higgs doublet, rather than just the 
gauge boson longitudinal ones, are considered Goldstone bosons and there is a light physical Higgs. 
In general, a physical Higgs boson in the spectrum would change our low-energy ($E < M_W$) description, 
due to the presence of flavour-violating sources in addition to the Yukawa couplings $Y_{u,d}$ 
(see for example Ref.~\cite{Giudice:2007fh}). However, there are specific limiting cases in which these 
extra sources do not contribute and all the flavour violation is encoded in $Y_{u,d}$. If for example 
the SM fermions couple bilinearly to the strong sector, these extra sources are universal parameters 
that can be simply re-absorbed. On the other side, if the SM fermions couple linearly to fermionic 
operators of the strong sector and therefore these couplings may provide extra flavour violation, 
our analysis applies only under the assumption of universal couplings.
Nevertheless, the models of this kind constructed up to now are not formulated as MFV models, as the flavour 
for the SM fermions is described by a weakly interacting sector, beside the strong one \cite{Redi:2011zi}. 
To extend them in this direction is beyond the scope of the present paper.

Finally, notice that we do not take in consideration in our analysis the possibility that the flavour breaking is non-linearly realized, as it has already been discussed in Ref.~\cite{Feldmann:2008ja,Kagan:2009bn}.

The structure of the manuscript is as follows. Sect. 2 describes the general formalism of the non-linear 
realization of the symmetry breaking mechanism and the effective MFV Lagrangian considered. In Sect. 3 
unitarity and CP violation are thoroughly discussed. Sect.~4 and Sect.~5 deal with the impact on $\Delta F=1$ 
and $\Delta F=2$ processes, respectively. The phenomenological impact is discussed in Sect.~6. Finally in 
Sect.~7 we conclude. Technical details are deferred to the Appendices.

%
\section{The MFV Lagrangian for a Heavy Higgs}
%

The standard implementation of the spontaneous SM electroweak symmetry breaking uses 
a complex scalar  field $\Phi$, which is a doublet under the $SU(2)_L$ gauge symmetry: 
\be
\Phi (x) \equiv 
\begin{pmatrix}
\varphi_+(x) \\ \varphi_0 (x)
\end{pmatrix} =
\frac{1}{\sqrt{2}} 
\begin{pmatrix}
\varphi_1 (x) - i \varphi_2 (x) \\
 v + H(x) + i \varphi_3 (x) \\
\end{pmatrix}\, ,
\ee
where $v=246$ GeV. A convenient way~\cite{Appelquist:1980vg,Longhitano:1980iz} to face the large $m_H$ 
limit is to represent the Higgs field through a  field $\MH$ transforming as a $(2,2)$ of the global 
$SU(2)_L\times SU(2)_R$ symmetry: 
\be
\MH (x) \equiv \sqrt{2} 
\begin{pmatrix}
\widetilde{\Phi} (x) & \Phi (x)
\end{pmatrix} = \sqrt{2} 
\begin{pmatrix}
\varphi_0^* (x) & \varphi_+ (x) \\
- \varphi_+^* (x)  & \varphi_0 (x) \\
\end{pmatrix}\,,
\ee
where $\widetilde\Phi = i \tau_2 \Phi^*$. Under the $SU(2)_L\times U(1)_Y$ gauge group the field $\MH$ 
transforms as
\be
\MH(x) \raw L(x) \, \MH(x) \, R^\dagger(x) \, ,
\ee 
with $L$ and $R$ the $SU(2)_L$ and $U(1)_Y$ gauge transformations, given respectively by
\be
L(x) = \exp \left\{ i\vec{\epsilon}_L(x) \cdot \frac{\vec{\tau}}{2} \right\}\,, \qquad \qquad 
R(x) = \exp \left\{ i\epsilon_Y (x) \frac{\tau_3}{2} \right\}  \, ,
\ee
where the  $\tau_3$ dependence reflects  the opposite hypercharges of $\Phi$ and $\widetilde \Phi$.
Accordingly, the Higgs scalar potential can be rewritten as:
\be
V(\MH) = \frac{1}{4}\,\lambda\,\left(\frac{1}{2}\,\tra\left[\MH^\dagger \MH\right]+
         \frac{\mu^2}{\lambda}\right)^2 \,.
\ee
For $\mu^2 < 0$, $\MH$ acquires a non-vanishing vev, 
\be
\langle \MH^\dagger\MH \rangle = v^2\, \mathbb{1}  \qquad {\rm with} \qquad 
v^2 = -\frac{\mu^2}{\lambda} \,,
\ee
and the physical Higgs field gets a mass $m_H=\sqrt{-2 \mu^2}$ breaking the global $SU(2)_L\times 
SU(2)_R$ symmetry down to $SU(2)_C$ and the local $SU(2)_L \times U(1)_Y$ gauge symmetry down to 
$U(1)_{EM}$. In the limit of large Higgs mass, the physical scalar degree of freedom can be 
formally decoupled sending $\mu ,\lambda \raw \infty$ while keeping $v$ finite. A non-linear realization 
of the gauge symmetry breaking is thus obtained subject to the constraint $\MH^\dagger\MH = 
\MH\MH^\dagger = v^2\,\mathbb{1}$. All remaining low-energy degrees of freedom of $\MH$ can be described 
through an adimensional unitary matrix $\UH$,
\be
\UH (x) = \MH(x)/v
\ee
and its covariant derivative, 
\be
{\cal D_\mu} \UH \equiv \partial_\mu \UH\,+\, \frac{i\,g}{2} \tau_i\, W^i_\mu\, \UH \, - 
\, \frac{i\,g'}{2}\UH\,\tau_3\, B_\mu\,.
\label{Ucovdev}
\ee
It is customary~\cite{Longhitano:1980tm,Appelquist:1984rr,Cvetic:1988ey}  to use the following combinations 
of $\UH$ and ${\cal D_\mu} \UH$, which transform covariantly under the SM gauge group: 
\beq
\begin{aligned}
&\TL  =  \UH\tau_3 \UH^{\dagger}\,, \qquad\qquad  &&\TL \, \raw L\, \TL L^\dagger \,,\\
&\VL_\mu =  ({\cal D}_\mu \UH)\UH^{\dagger}\,, \qquad \qquad &&\VL_\mu \raw L\, \VL_\mu L^\dagger  \, .
\end{aligned}
\eeq
The power series in chiral dimensions is then a Taylor expansion whose terms are operators constructed out 
of $\TL$, $\VL_\mu/{v}$, $W_{\mu\nu}/{v}^2$ and $B_{\mu\nu}/{v}^2$.  

The Lagrangian describing the interaction between the gauge fields and the scalar sector reads:
\be
\mathcal{L} = - \frac{1}{4} W^i_{\mu\nu} W^{\mu\nu}_i - \frac{1}{4} B_{\mu\nu} B^{\mu\nu} + 
\frac{v^2}{4}\text{Tr}[\VL_\mu \VL^{\mu}]\,+\, \mathcal{\delta L}\,,
\ee
in which the first two terms have chiral dimensions $d_\chi=4$ and the third one has instead $d_\chi=2$. 
Those three terms preserve the global $SU(2)_L\times SU(2)_R$ symmetry.  $\mathcal{\delta L}$ contains other gauge 
invariant terms, but includes among them terms breaking the custodial symmetry that the quantum corrections 
inevitably induce. At lowest order $d_\chi =2$ there is only one term,
\be
 \mathcal{\delta L}_{d_\chi=2}= a_{WZ}\,\frac{v^2}{4}\,\tra[ \TL\,\VL_{\mu}]\,\text{Tr}[ \TL\,\VL^{\mu}]\,,
 \label{Lagrangiand=2}
\ee
which breaks the custodial symmetry inducing a shift in the $Z$-boson mass with respect to the $W$-boson 
mass. This coupling tends to be unacceptably large in naive models of a strong interacting Higgs sector, 
from the original technicolor formulation \cite{Susskind:1978ms,Dimopoulos:1979es} to its modern variants,  
if not opportunely protected by some additional custodial symmetry. 
Quantitatively, realistic models need to limit the intensity of this $\mathcal{\delta L}_{d_\chi=2}$ 
induced coupling to $a_{WZ} <0.001$, as it is well-known. 

A large number of operators appear at dimension $d_\chi=4$. Thirteen of them contribute to the 
so called oblique sector~\cite{Longhitano:1980tm}. None of them break the custodial symmetry, but 
are severely constrained by EW precision tests~\cite{Alcaraz:2009jr}. However they carry no flavour 
and in consequence they will not be further discussed below. In the following, instead, we will 
explicitly concentrate on those operators -among the $d_\chi=4$ leading ones- relevant for flavour 
changing transitions. 

In the non-linear formalism, the Yukawa interactions in Eqs.~(\ref{YukawasL}) may be written as:
\be
\LL_Y= \frac{v}{\sqrt{2}}\,\ov{Q}_L \,\mathcal{Y}\, \UH \,Q_R + \hc \,,
\label{YukawasNL}
\ee
where $Q_R=(u_R,d_R)$ and $\mathcal{Y}$ is a $2\times 2$ block matrix built up from the up and down 
Yukawas matrices. We will work in the usual MFV basis in which the down-sector Yukawa couplings 
are taken to be diagonal in flavour, with all flavour violation parameterized via the up sector, 
\be
\mathcal{Y} \equiv
\begin{pmatrix}
Y_U & 0 \\
0   & Y_D\\
\end{pmatrix} =
\begin{pmatrix}
V^\dagger \mathbf{y}_U & 0 \\
0   & \mathbf{y}_D \\
\end{pmatrix} \, ,
\ee
where $\mathbf{y}_U$, $\mathbf{y}_D$ are diagonal matrices whose elements are the Yukawa eigenvalues 
and $V$ denotes the CKM matrix. It is also convenient and customary to define the combination:
\be
\lambda_{F} \equiv Y_U Y_U^\dagger + Y_D Y_D^\dagger = V^\dagger \mathbf{y}_U^2 V +  \mathbf{y}_D^2 \,,
\label{lambdaFC}
\ee
which transforms as a $(8,1,1)$ under the flavour group $\mathcal{G}_f$ and will determine the strength 
of new flavour effects. In practice, as the SM Yukawa couplings for all fermions except the top are 
small\footnote{In scenarios in which the bottom Yukawa coupling may be of order one, this assumption 
has to be accordingly modified.}, the only relevant non-diagonal flavour structure will be given by the 
contraction of two $Y_U$ with $(\lambda_{F})_{ij} \equiv (Y_U Y_U^\dagger)_{ij} \approx \mathbf{y}^2_t 
V^*_{ti}V_{tj}$, for $i\neq j$.

\boldmath
\subsection{$d_{\chi}=4$ Contributions}
\unboldmath

Consider the $d_{\chi}=4$ part of the Lagrangian which contains only combinations of the field $\UH$ 
and fermions:
\be
\mathcal{\delta L}_{d_\chi=4  }\,=\, a_i\, \mathcal{O}_i \,.
\ee
Operators containing two right-handed fields $Q_R$ will be flavour conserving at leading order in the 
spurion expansion\footnote{With more insertions of the spurions, it may also result in flavour non-diagonal 
operators, but they will be Yukawa suppressed and in this sense subdominant.} and of no interest in 
what follows. Four independent $d_{\chi}=4$ 
flavour violating operators involving left-handed quarks $Q_L$ and the $\UH$ field can, instead, 
be written~\cite{Appelquist:1984rr,Cvetic:1988ey,Espriu:2000fq}: 
\begin{align}
&\mathcal{O}_{1} = \frac{i}{2}\bar{Q}_L\lambda_{F} \gamma^{\mu}\left\{\TL , \VL_{\mu} \right\} Q_L\,,\qquad
&&\mathcal{O}_{2} = i\bar{Q}_L\lambda_{F}\gamma^{\mu}\VL_{\mu} Q_L\,, 
\label{siblings}\\
&\mathcal{O}_{3} = i\bar{Q}_L\lambda_{F}\gamma^{\mu}\TL \VL_{\mu} \TL Q_L\,,\qquad
&&\mathcal{O}_{4} = \frac{1}{2}\bar{Q}_L\lambda_{F}\gamma^{\mu}\left[\TL , \VL_{\mu}\right] Q_L \,.
\label{nosiblings}
\end{align}
The operator $\mathcal{O}_{2}$ has been  previously singled out in the context of composite Higgs models, 
see for example Ref.~\cite{Agashe:2006at}.
Notice from Eqs.~(\ref{siblings}) and (\ref{nosiblings}) the interesting fact that in the non-linear 
realization of MFV a CP-odd operator, $\cO_{4}$, emerges at leading order. Much like in the strong 
CP problem, the \emph{mere presence} of $\cO_{4}$ will violate CP without the need of introducing 
complex coefficients, as done for example in Refs.~\cite{Mercolli:2009ns,Ellis:2010xm}. 
The presence of this operator is a slight modification of the MFV ansatz\footnote{Our only requirement 
is the  invariance under the flavour group $\Gf$ for all operators built out of the SM model fields 
(and $\UH$) and the spurions $Y_{U,D}$. In Ref.~\cite{D'Ambrosio:2002ex} the invariance under CP was 
additionally assumed by restraining all operator coefficients to be real; no genuine CP-odd operator 
stems at leading order  of the linear expansion (the sibling of $O_{4}$ would appear in it only at 
higher order). In our approach we will keep the new source of CP violation naturally present at leading 
order only for the non-linear expansion, for its theoretical and phenomenological interest.} as defined  
in Ref.~\cite{D'Ambrosio:2002ex}. We will analyze in detail its implications  in the following sections.

\subsection{Relation to the Linear Expansion}

In the linear realization of the Higgs mechanism, the leading MFV ``flavour" operators appear at dimension 
$d=6$. In that case, four operators involving the Higgs field and two fermions were found; two of those, 
dubbed $\mathcal{O}_{H_1}$ and $\mathcal{O}_{H_2}$ in Ref.~\cite{D'Ambrosio:2002ex}, produce the same 
low-energy effects than our operators $\mathcal{O}_{1}$ and $\mathcal{O}_{2}$ in Eq.~(\ref{siblings}) 
as can be seen replacing in the former the Higgs field by its vev. The {\it siblings} of the other two 
(linear) operators, labeled as $\mathcal{O}_{G_1}$ and $\mathcal{O}_{F_1}$ in Ref.~\cite{D'Ambrosio:2002ex} 
do not appear at the leading $d_{\chi}=4$ order in the non-linear expansion, but only at $d_{\chi}=5$. 
Conversely, the linear siblings of our leading operators $\mathcal{O}_{3}$ and $\mathcal{O}_{4}$ in 
Eq.~(\ref{nosiblings}) have not been considered in Ref.~\cite{D'Ambrosio:2002ex}: they would have 
dimension $d=8$ in the linear realization, as it can be easily checked with the correspondences in 
Eqs.~(\ref{NLtoL1})--(\ref{NLtoL3}) in Appendix 1. In consequence, the phenomenological signals of 
MFV is expected to exhibit notable differences between the two scenarios.

\boldmath
\subsection{The Effective Low-Energy Lagrangian}
\unboldmath

Aside from fermion masses, three parameters are relevant to our discussions: the weak isospin coupling 
constant $g$, the hypercharge coupling constant $g'$, and EW symmetry breaking scale $v$. The impact 
on them of the operators introduced in Eq.~(\ref{siblings}) and (\ref{nosiblings}) is best seen looking  
at the modification of the low-energy effective Lagrangian in the unitary gauge, i.e. $\UH=\mathbb{1}$:
\bea
\hspace{-1cm} \mathcal{\delta L}_{d_\chi=4} &=&
 - \frac{g}{\sqrt{2}}\left[ W^{\mu+}\bar{U}_L \gamma_\mu (a_W+ia_{CP}) \left(\mathbf{y}_U^2 V + 
   V\mathbf{y}_D^2\right) D_L + h.c. \right]+ \nn \\
 & & -\frac{g}{2c_W}Z^\mu\left[
 a^{u}_Z\bar{U}_{L}\gamma_\mu \left(\mathbf{y}_U^2+V\mathbf{y}_D^2 V^\dagger\right) U_L+
 a_Z^d\bar{D}_{L}\gamma_\mu \left(\mathbf{y}_D^2+V^\dagger\mathbf{y}_U^2 V\right) D_L\right]
\label{DevSM}
\eea
where $c_W$ and $s_W$ stand for the cosine and sine of the weak angle $\theta_W$. The new couplings 
are codified through 
\begin{equation}
\begin{array}{ll}
a_Z^u\equiv a_1+a_2+a_3\,, \qquad \quad &
a_Z^d\equiv a_1-a_2-a_3\,,\,\,\,\\
a_W\equiv a_2-a_3\,,\,\,\, &
a_{CP}\equiv - a_4\,.\,\,\,
\end{array}
\label{ChangeBasisZW}
\end{equation}

It is customary to fix the values of $g$, $g'$ and $v$ by taking the three best known experimental 
quantities~\cite{Nakamura:2010zzi}: the fine structure constant $\alpha$ -as determined from the quantum 
Hall effect, the Fermi constant $G_F$ -as extracted from the muon lifetime- and the $Z$-boson mass $M_Z$ 
as determined from the Z line-shape scan at LEP 1. Eqs.~(\ref{DevSM}) and (\ref{ChangeBasisZW}) show that 
none of the operators in Eqs.~(\ref{siblings}) and (\ref{nosiblings}) contribute to $\alpha$ or $M_Z$, 
while three of them {\it a priori} correct $G_F$: that are $\mathcal{O}_2$, $\mathcal{O}_3$ and 
$\mathcal{O}_4$. Nevertheless, as the determination of $G_F$ only involves the first family of fermions, 
the impact of the new couplings discussed here can be neglected within the MFV scheme, where the effects 
of new operators come with a proportionality to the Yukawa couplings of the fermions involved. 

On the contrary, those four operators of Eqs.~(\ref{siblings}) and (\ref{nosiblings}) will have 
important phenomenological consequences for transitions involving heavier quarks. The contributions 
of $\mathcal{O}_2$, $\mathcal{O}_3$ and $\mathcal{O}_4$ will induce non-unitarity quark mixing and 
new CP-odd effects to fermion-$W$ couplings. $\mathcal{O}_1$ affects instead fermion-$Z$ couplings, 
to which $\mathcal{O}_2$ and $\mathcal{O}_3$ also contribute, resulting in tree-level flavour changing 
neutral currents (FCNC).

\section{Non Unitarity and CP Violation}

The tree-level modifications of the $W$-fermion couplings from $\mathcal{O}_2$, $\mathcal{O}_3$ and 
$\mathcal{O}_4$ can be parameterized as a modified, non-unitary, mixing matrix:
\begin{equation}
\tilde{V}_{ij}=V_{ij}\left[1+(a_W+ia_{CP})(y_{u_i}^2+y_{d_j}^2)\right] \,.
\label{Vtilde}
\end{equation}
In the limit of vanishing Yukawa couplings but the top one\footnote{These relations should be modified 
if one works in a framework in which $y_b \approx y_t$.}, a good estimation of the departures from 
unitarity can be read from:
\bea
\sum_k\tilde{V}_{ik}^*\tilde{V}_{jk} & \simeq & \delta_{ij} + 
  \left[2 \, a_W \,y_t^2 + (a_W^2 + a_{CP}^2) \, y_t^4\right]\delta_{it} \delta_{jt}\,,\label{lambda} \\
\sum_k\tilde{V}_{ki}^*\tilde{V}_{kj} &\simeq & \delta_{ij}+
 \left[2\,a_W\,y_t^2 + (a_W^2 + a_{CP}^2)\, y_t^4\right] V_{ti}^* V_{tj}\,. \label{lambdap}
\eea
As expected from Eq.~(\ref{lambdaFC}), sizable unitarity corrections are expected in transitions 
involving the third family of quarks, still not severely constrained by experiments. In contrast, 
the unitarity corrections induced on the first two family sectors, being suppressed by small Yukawa 
eigenvalues and small mixing angles, are $\mathcal{O}(10^{-4})$, perfectly in agreement with present  
bounds~\cite{Nakamura:2010zzi}. Eqs.~(\ref{lambda}) and (\ref{lambdap}) illustrate in addition that 
$a_{CP}$-dependent non-unitarity corrections appear only at second order. 
 
Let us consider now CP violation. The first remarkable aspect is that the new source of CP-violation 
only requires two fermion families, contrary to the SM case, see Eq.~(\ref{Vtilde}). This is evident 
from the non-unitary character of the NP correction. Indeed, even for two families the $a_{CP}$-dependent 
contributions cannot be rotated away through fermionic-field re-definitions due to the non-trivial 
flavour structure of both  the up and down quark sectors, when the mixing matrix is non-unitary.
This translates, for instance, in that the usual parametrization-invariant definition of the angles 
of the unitarity triangles now reads:
\begin{equation} 
\begin{split}
\mbox{arg}\left(-\frac{\tilde{V}^*_{ik} \tilde{V}_{il}}{\tilde{V}^*_{jk}\tilde{V}_{jl}}\right) = & \,\,
\mbox{arg}\left(-\frac{V^*_{ik} V_{il}}{V_{jk}^*V_{jl}}\right) 
+ a_{CP}\,\big[2\,a_W\,\left(y_{u_j}^2-y_{u_i}^2\right)\left(y_{d_l}^2-y_{d_k}^2\right) + \\ & 
\hskip-2cm
- \left(3\,a_W^2-a_{CP}^2\right)\left(y_{u_j}^2-y_{u_i}^2\right)\left(y_{d_l}^2-y_{d_k}^2\right)
\left(y_{u_i}^2+y_{u_j}^2+y_{d_k}^2+y_{d_l}^2\right)\big]
+\mbox{O}(a^4)\label{CPphaseGeneral} \,,
\end{split}
\end{equation}
showing that:
\begin{itemize}
\item[-]
All corrections are proportional to $a_{CP}$, as expected from the fact that the SM source 
of CP-violation  is the only  one remaining in the absence of $\mathcal{O}_4$;
\item[-]
Two quark families, non-degenerate in both the up and down sectors, are necessary and 
sufficient to induce physical CP-odd effects proportional to $a_{CP}$. In particular, no CP-odd effects are present in the one-family case;
\item[-]
New CP-odd contributions appear generically at quadratic order $O(a_{CP}a_W)$. In contrast, 
when the operator $\cO_4$ is considered by itself (e.g. $a_W=a^u_Z=a^d_Z=0$ in Eq.~(\ref{DevSM}))
the leading correction is $O(a_{CP}^3)$ or higher.
\end{itemize}
Overall, the arguments above imply that the leading CP-odd effects should be $\sim a^2_X y_t^2y_b^2$, 
when two operators are considered simultaneously. In particular, using Eq.~(\ref{CPphaseGeneral}), one can 
give an estimation of how the angles of the $B_d$ unitarity triangle, $(\alpha,\beta,\gamma)$ get 
modified at leading order: 
\beq
\begin{aligned}
&\mbox{arg}\left(-\frac{\tilde{V}^*_{tb} \tilde{V}_{td}}{\tilde{V}^*_{ub}\tilde{V}_{ud}}\right)
\simeq \alpha + 2\,y_b^2\,y_t^2\,a_W\,\,a_{CP}\,,\\
&\mbox{arg}\left(-\frac{\tilde{V}^*_{cb} \tilde{V}_{cd}}{\tilde{V}^*_{tb}\tilde{V}_{td}}\right)
\simeq \beta - 2\,y_b^2\,y_t^2\,a_W\,a_{CP}\,,\\
&\mbox{arg}\left(-\frac{\tilde{V}^*_{ub} \tilde{V}_{ud}}{\tilde{V}^*_{cb}\tilde{V}_{cd}}\right)
\simeq \gamma - 2\,y_c^2\,y_b^2\,a_W\,a_{CP} \simeq \gamma \,.
\label{angles}
\end{aligned}
\eeq
Interestingly, the angle $\gamma$ remains  free from new physics contamination, up to corrections 
of order $y_c^2/y_t^2$ relatively to those for $\alpha$ and $\beta$.

In the next sections we will discuss $\Delta F=1$ and $\Delta F=2$ transitions. 

\boldmath
\section{$\Delta F=1$ Observables}
\label{sec:DeltaF1}
\unboldmath

\subsection*{FCNC}
The $d_\chi=4$ operators $\mathcal{O}_{1}$, $\mathcal{O}_{2}$ and $\mathcal{O}_{3}$ induce tree-level 
FCNC processes, as can be seen from the $Z$ couplings of the effective Lagrangian of Eq.~(\ref{DevSM}). 
This results in important constraints on their coefficients. 
The siblings of $\mathcal{O}_{1}$ and $\mathcal{O}_{2}$ for the linear regime (i.e. $\mathcal{O}_{H1}$ 
and $\mathcal{O}_{H2}$) have been thoroughly analyzed in \cite{D'Ambrosio:2002ex}. The bounds obtained 
there can be straightforwardly applied also to our case to constrain the specific combinations of 
coefficients that modify the neutral current.

For the purposes of the present work, only $\Delta F=1$ processes involving $K$ and $B$ mesons need 
to be considered. In fact, as can be seen in Eq.~(\ref{DevSM}) the new up-type tree-level FCNC 
contributions are governed by the down-type spurion $V\mathbf{y}_D^2 V^\dagger$, subleading with respect 
to $\lambda_{F}$, by at least a factor $y^2_b/y^2_t$. Therefore only bounds on $a_Z^d$ are 
going to be discussed in the following. 

The down-type low-energy FCNC effective Lagrangian is usually written as
\begin{equation}
\frac{G_F\alpha}{2\sqrt{2}\pi s_W^2}V_{ti}^*V_{tj}\sum_n C_n \, \mathcal{Q}_n+\mbox{h.c.}\,,
\label{FCNCbasisSM}
\end{equation}
where the Wilson coefficient $C_n$ contain both the SM and the NP contributions:
\be
C_n=C^{SM}_n+C^{NP}_n
\ee  
and $\mathcal{Q}_n$ stands for the FCNC operators constructed with SM fields, the basis of which is 
customarily taken as:
\begin{equation}
\begin{array}{ll}
\mathcal{Q}_{\bar\nu\nu}= \bar{d}_i\gamma_\mu(1-\gamma_5)d_j\,\bar{\nu}\gamma^\mu(1-\gamma_5)\nu \,,\qquad
& \mathcal{Q}_{7}=e_q\,\bar{d}_i\gamma_\mu(1-\gamma_5)d_j\,\bar{q}\gamma^\mu(1+\gamma_5) q \,, \\
\mathcal{Q}_{9V}=\bar{d}_i\gamma_\mu(1-\gamma_5)d_j\,\bar{\ell}\gamma^\mu \ell\,,
& \mathcal{Q}_{9}=e_q\,\bar{d}_i\gamma_\mu(1-\gamma_5)d_j\,\bar{q}\gamma^\mu(1-\gamma_5) q \,,\\
\mathcal{Q}_{10A}=\bar{d}_i\gamma_\mu(1-\gamma_5)d_j\,\bar{\ell}\gamma^\mu\gamma_ 5 \ell\,,
& \mathcal{Q}_{7\gamma}=\frac{m_i}{g^2}\bar{d}_i(1-\gamma_5)\sigma_{\mu\nu}d_j\,(e\,F^{\mu\nu}) \,,\\
& \mathcal{Q}_{8G}=\frac{m_i}{g^2}\bar{d}_i(1-\gamma_5)\sigma_{\mu\nu}T^ad_j\,(g_s\,G_a^{\mu\nu})\,. 
\end{array}
\end{equation}
A sum over all quark species is understood in $\mathcal{Q}_{7,9}$, with $e_q$ denoting the quark 
electric charge. The main SM contribution to the Wilson coefficients results from top quark loop, 
while the effects of our $d_\chi=4$ operators appear already at tree-level. The leading NP 
contributions coming from the non-linear MFV operators read:
\be
\begin{array}{lll}
C^{NP}_{\nu\bar\nu}= -\kappa \, y_t^2 \,a_Z^d\,,  &\qquad & 
C^{NP}_{7} = +2\kappa \, s_W^2 \, y_t^2 \,a_Z^d\,,  \\
C^{NP}_{9V}= \kappa\,(1-4s_W^2) \, y_t^2 \,a_Z^d \,,
&\qquad \qquad & 
C^{NP}_{9} = - 2 \kappa\, c_W^2 \, y_t^2 \, a_Z^d\,,  \\
C^{NP}_{10A}= - \kappa \, y_t^2 \, a_Z^d\,,
&\qquad &   
C^{NP}_{7\gamma}=C^{NP}_{8G}=0\, .\\
\end{array}
\ee
where $\kappa \equiv \pi s^2_W /(2\,\alpha)$ reflects the relative strength of the NP tree-level 
contribution  with respect to the loop-suppressed SM one. Note that the CP-odd operator $\mathcal{O}_4$ 
does not modify the $Z$ couplings and in consequence it has no impact at this order.

Different rare decays of mesons can be analyzed~\cite{Hurth:2008jc} in order to bound the $a_Z^d$ 
coefficient. The cleanest channels -those with less hadronic uncertainties- are reported in Table~\ref{tab1} 
and the following overall constraint $-0.044 < a_Z^d < 0.009$ can be extracted, at 95\% of CL.

\begin{table}[t]
\centering
\begin{tabular}{c c c}
Operator & Observable & Bound (@ 95\% C.L.)\\
\hline
$\mathcal{O}_{9V}$ & $B\rightarrow X_s \emph{l}^+ \emph{l}^-$& $-0.811<a_Z^d<0.232$\\
$\mathcal{O}_{10A}$ & $B\rightarrow X_s \emph{l}^+ \emph{l}^-$
,$B\rightarrow \mu^+ \mu^-$& $-0.050<a_Z^d<0.009$\\
$\mathcal{O}_{\bar\nu\nu}$ & $K^+\rightarrow \pi^+ \bar\nu \nu$
& $-0.044<a_Z^d<0.133$\\
\end{tabular}
\caption{\it  FCNC bounds~\cite{Hurth:2008jc} on the combination of operator coefficients
$a_Z^d$, obtained from a tree-level analysis.}
\label{tab1}
\end{table}

The bounds on $\Delta F=1$ FCNC transitions among up-type quarks, $a_Z^u$, can be easily predicted to be  
of $\mathcal{O}(a_Z^d\,y_b^2/y_t^2)$ and consequently do not provide any interesting additional information.

\mathversion{bold}
\subsection*{$B^+\to\tau^+\nu$}
\mathversion{normal}

The  branching ratio for $B^+\to\tau^+\nu$ will be of particular relevance in the  phenomenological 
analysis below, and therefore we consider its impact in detail. In the SM, this decay occurs as a 
tree-level charged current process. The main NP correction enters through the modification of the 
CKM matrix element $\tilde V_{ub}$, see Eq.~(\ref{Vtilde}), resulting in:
\beq
BR(B^+\to\tau^+\nu)=\dfrac{G_F^2\,m_{B^+}\,m_\tau^2}{8\pi}\left(1-\dfrac{m_\tau^2}{m^2_{B^+}}\right)^2
\,F^2_{B^+}\,| V_{ub}|^2\,\left|1+\left(a_W+i\,a_{CP}\right)\,y_b^2\right|^2\,\tau_{B^+}\,,
\label{BRtau}
\eeq 
with $F_{B^+}$ the $B$ decay constant\footnote{The SM lepton-$W$ couplings have been assumed in writing 
Eq.~(\ref{BRtau}). Even if we are not considering the lepton sector in our scenario, those couplings 
are strongly constrained by the SM electroweak analysis and therefore any analogous NP modification 
in the lepton sector should be safely negligible.}. FCNC $Z$-mediated contributions to this process 
appear at the one-loop level and can be safely neglected. 

\mathversion{bold}
\section{$\Delta F=2$ Observables and $B$ Semileptonic CP-Asymmetry}
\mathversion{normal}
\label{sec:DeltaF2SemileptonicAsym}

In this section we provide analytical expressions for the NP contributions to $K$ and $B$ mesons 
oscillations and $B$ semileptonic CP-asymmetry in terms of the three parameters $a_W$, $a_{CP}$ and 
$a_Z^d$ as there is no dependence on $a_Z^u$ in the observables considered. The formula reported here, 
concerning the NP part, have been obtained keeping only the dominant term in $\lambda_{F}$, 
Eq.~(\ref{lambdaFC}), neglecting the Yukawa couplings of the lightest quarks~\footnote{When considering 
the $B$ system in the formulae below, only terms proportional to $y_t^2$ and $y_b^2$ will be described. 
For $K$ mixing instead, $y_c$-dependent terms will be also retained.} and retaining solely the 
leading contributions in each of the new parameters. In practice this means that the expressions below 
will only contain terms at most quadratic in $a_W$, $a_{CP}$, $a_Z^d$ or their combinations. A detailed 
discussion on the Wilson coefficients and the renormalization group QCD evolution is deferred to 
App.~\ref{App:PhemSection}.

\mathversion{bold}
\subsection{$\Delta F=2$ Observables}
\mathversion{normal}
\label{sec:DeltaF2}

The  modified $W$ and $Z$ couplings exhibited by the effective low-energy Lagrangian in Eq.~(\ref{DevSM}) 
induce deviations from the SM predictions for $\Delta F=2$ observables. The main corrections enter through 
the box diagrams with $W$-bosons exchange. However contributions from tree-level FCNC $Z$ diagrams can 
be relevant as well and will be considered, while $Z$-mediated boxes and weak penguin diagrams can be 
safely neglected being suppressed with respect to the tree-level $Z$ contributions.

The effective Hamiltonian for $\Delta F=2$ transitions is usually written as:
\beq
\cH_\text{eff}^{\Delta F=2} =\dfrac{G_F^2\,M^2_{W}}{4\pi^2}\, C(\mu)\,Q\,,
\label{Heff-general}
\eeq
where $Q$ is the operator describing the neutral meson mixing:
\beq
Q=(\bar d_i^\alpha\,\gamma_\mu\,P_L\,d_j^\alpha)(\bar d_i^\beta\,\gamma^\mu\,P_L\,d_j^\beta) \,, 
\label{EffectiveOperatorMesonOscillations}
\eeq
and $C(\mu)$ denote the Wilson coefficient evaluated at a scale $\mu$. The mixing amplitude $M^i_{12}$ 
($i=K,d,s$) is defined from the effective hamiltonian by:
\beq
M^K_{12}=\dfrac{\langle \bar K^0| \cH_\text{eff}^{\Delta S=2}|K^0\rangle^*}{2\,m_K}\,,\qquad\qquad
M^q_{12}=\dfrac{\langle \bar B_q^0| \cH_\text{eff}^{\Delta B=2}|B_q^0\rangle^*}{2\,m_{B_q}}\,,
\eeq
with $q=d,\,s$. For the $K$ system, the mixing amplitude can be written as the sum of the SM and the NP 
contributions\footnote{The expression for $M^K_{12}$ is phase-convention dependent. In the following 
we will give all the results in the convention in which the phase of the $K\to\pi\pi$ decay amplitude 
is vanishing.} $M^K_{12}=(M^K_{12})_{SM}+(M^K_{12})_{NP}$.  Neglecting all contributions proportional 
to $y_{u,d,s}$ and $y_c^n$ with $n>2$ one has:
\bea
(M^K_{12})_{SM} &=& R_{K}\Big[\eta_2\,\la_t^2\,S_0(x_t)+\eta_1\,\la_c^2\,S_0(x_c)+
   2\,\eta_3\,\la_t\,\la_c\,S_0(x_c,\,x_t)\Big]^*\,, \nn \\
(M_{12}^K)_{NP} &=& R_{K}\,\Bigg[\eta_2\,\lambda_t^2\,\left(y_t^2\,(2\,a_W\,+y_t^2\,a^2_{CP})\,G(x_t)+
   \dfrac{(4\,\pi\,y_t^2\,a_Z^d)^2}{g^2}\right)+2\eta_1\,\lambda_c^2\,a_W\,y_c^2\,G(x_c)+ \nn \\
&&\quad\quad +2\,\eta_3\,\lambda_t\,\lambda_c\,\left(y_t^2\,(2\,a_W\,+a^2_{CP}\,y_t^2)\,H(x_t,x_c)+
   2\,a_W\,y_c^2\,H(x_c,x_t)\right)\Bigg]^*\,,
\label{M12NP}
\eea
where $\eta_i$ denote terms due to QCD higher order effects, $S_0$, $G$ and $H$ are loop functions 
defined in App.~\ref{App:PhemSection} and 
\beq
R_K\equiv\dfrac{G_F^2\,M_W^2}{12\,\pi^2}F_K^2\,m_K\,\hat B_K\,,
\eeq
with $\hat B_K$ the scale-independent hadronic B-mixing matrix element~\cite{Laiho:2009eu}. 
Eq.~(\ref{M12NP}) shows only the leading terms in each parameter: linear terms in $a_W$ and quadratic 
terms in $a_{CP}$ and $a_Z^d$. Indeed, the effects of non-linear terms in $a_W$ turn out to be 
negligible in the numerical analysis of the next section. 

The $K_L-K_S$ mass difference and the CP-violating parameter $\vep_K$ are given by
\beq
\begin{aligned}
&\Delta M_K=2\,\Big[\re(M^K_{12})_{SM}+\re(M^K_{12})_{NP}\Big]\,,\\
&\varepsilon_K=\dfrac{\kappa_\epsilon\, e^{i\,\varphi_\epsilon}}{\sqrt{2}\,(\Delta M_K)_\text{exp}}
  \Big[\im\left(M_{12}^K\right)_{SM}+\im\left(M_{12}^K\right)_{NP}\Big]\,,
\end{aligned}
\eeq
where $\varphi_\epsilon$ and $\kappa_\epsilon$ (see Table~\ref{tab:input}) 
account for $\varphi_\epsilon\neq\pi/4$ and include long-distance contributions to $\im (\Gamma_{12})$ 
and $\im (M_{12})$.

For the $B_{d,s}$ systems, it is useful to move to a slightly different notation for the mixing amplitude 
$M_{12}^q$:\footnote{The expression for $M_{12}^q$ is phase-convention dependent and we adopt the convention 
in which the  decay amplitudes of the corresponding processes, $B_d^0\to\psi K_S$ and $B_s^0\to \psi\phi$, 
have a vanishing phase.}
\beq
M_{12}^q=(M_{12}^q)_\text{SM}\,\mathcal{C}_{B_q}\,e^{2\,i\,\varphi_{B_q}}\,,
\label{NotationM12withNP}
\eeq
where $\mathcal{C}_{B_{d,s}}$ and $\varphi_{B_{d,s}}$ parametrize the NP effects, while the SM contribution 
is given by:
\beq
M_{12}^q=R_{B_q}\,\left[\la_t^2\,S_0(x_t)\right]^*\,,\qquad\text{with}\qquad
R_{B_q}\equiv \dfrac{G_F^2\,M_W^2}{12\,\pi^2}F_{B_q}^2\,m_{B_q}\,\hat B_{B_q}\,\eta_B\,,
\eeq
with $F_{B_q}$ and $\hat B_K$ denoting the neutral $B$ decay constant and mixing hadronic matrix elements, 
respectively. The mass differences in the $B_{d,s}$ systems are given by
\beq
\Delta M_q=2\,|M_{12}^q|\equiv(\Delta M_q)_\text{SM}\,\mathcal{C}_{B_q}\,,
\eeq
with
\beq
\mathcal{C}_{B_d}=\mathcal{C}_{B_s}=\Bigg|1+
2\,a_W\,\left(y_t^2\,\dfrac{G(x_t)}{S_0(x_t)}+y_b^2\right)+\dfrac{(4\,\pi\, y_t^2\,a_Z^d)^2}{g^2\,S_0(x_t)}+
2\,i\,a_W\,a_{CP}\,y_t^2\,y_b^2\,\dfrac{G(x_t)}{S_0(x_t)}\Bigg|\,.
\label{CBds}
\eeq 
The mixing-induced CP asymmetries $S_{\psi K_S}$ and $S_{\psi\phi} $ in the decays $B_d^0\to \psi\,K_S$ 
and $B_s^0\to\psi\,\phi$, respectively, are described by
\beq
 S_{\psi K_S} =\sin(2\,\beta+2\,\varphi_{B_d})\,,\qquad\qquad
 S_{\psi\phi} =\sin(2\,\beta_s-2\,\varphi_{B_s})\,,
\eeq
where $\beta$ and $\beta_s$ are angles in the unitary triangles,
\beq
\beta\equiv\arg\left(-\dfrac{V_{cb}^*\,V_{cd}}{V_{tb}^*\,V_{td}}\right)\,,\qquad\qquad
\beta_s\equiv\arg\left(-\dfrac{V_{tb}^*\,V_{ts}}{V_{cb}^*\,V_{cs}}\right)\,,
\eeq
and the new phases are given by
\beq
\varphi_{B_d}=\varphi_{B_s}=2\,a_W\,a_{CP}\,y_t^2\,y_b^2\dfrac{G(x_t)}{S_0(x_t)}\,.
\label{NPphaseB}
\eeq

It is interesting to point out that, at this level of approximation, the NP contributions to 
$\Delta M_{B_d}$ and $\Delta M_{B_s}$ are equal (see Eq.~(\ref{CBds})) and as a result the ratio 
$R_{\Delta M_B}\equiv\Delta M_{B_d}/\Delta M_{B_s}$ turns out to be not affected by NP. Any deviation 
from the SM value of this observable is then negligible in our framework. Another very clean observable 
is the ratio between $\Delta M_{B_d}$ and the $B^+\to\tau^+\nu$ branching ratio \cite{Isidori:2006pk}:
\beq
R_{BR/\Delta M}=\dfrac{3\,\pi\,\tau_{B^+}}{4\,\eta_B\,\hat B_{B_d}\,S_0(x_t)}
\dfrac{m^2_\tau}{M_W^2}\dfrac{\left|V_{ub}\right|^2}{\left|V^*_{tb}\,V_{td}\right|^2}
\left(1-\dfrac{m_\tau^2}{m_{B_d}^2}\right)^2\,\dfrac{\left|1+\left(a_W+i\,a_{CP}\right)\,
y_b^2\right|^2}{\mathcal{C}_{B_d}}\,,
\label{R_BRoverDM}
\eeq 
where we took $m_{B^+}\approx m_{B_d}$, well justified considering the errors in the other quantities 
in this formula.

Finally, notice that the NP contributions to $\Delta M_q$ ($q=K,\,d,\,s$) and $\varepsilon_K$ are 
all proportional to $y_t^2$ and therefore one may expect large effects on these observables, driven by 
$a_W$, $a_{CP}$, and $a_Z^d$. In contrast, the new contributions to $S_{\psi K_S}$ and $S_{\psi\phi}$ 
appear to be proportional to $y_b^2$ and only small deviations from their SM values are thus expected. 

\mathversion{bold}
\subsection{The $B$ Semileptonic CP-Asymmetry}
\mathversion{normal}
\label{sec:SemileptonicAsym}

In the $B_q$ systems, in addition to $\Delta M_q$, $S_{\psi K_S} $ and $S_{\psi\phi}$, a fourth 
observable provides rich information on  meson mixing: the like-sign dimuon charge asymmetry of 
semileptonic $b$ decays $A^b_{sl}$: 
\beq
A^b_{sl}\equiv \dfrac{N_b^{++}-N_b^{--}}{N_b^{++}+N_b^{--}}\,,
\eeq
where $N_b^{++}$ and $N_b^{--}$ denote the number of events containing two positively or negatively 
charged muons, respectively. In $p\bar p$ colliders, such events can only arise through $B_d^0-\bar 
B_d^0$ or $B_s^0-\bar B_s^0$ mixings. Due to the intimate link with meson oscillations, $A^b_{sl}$ 
is also called semileptonic CP-asymmetry and gets contributions from both $B_d$ and $B_s$ systems 
\cite{Abazov:2011yk,Lenz:2011ww}:
\beq
A^b_{sl}=(0.594\pm0.022)\,a^d_{sl}+(0.406\pm0.022)\,a^s_{sl}\,,
\eeq
where
\beq
\begin{aligned}
&a^d_{sl}\equiv\left|\dfrac{\left(\Gamma_{12}^d\right)_{SM}}{\left(M_{12}^d\right)_{SM}}
\right|\sin\phi_d=(5.4\pm1.0)\times10^{-3}\,\sin\phi_d\,,\\
&a^s_{sl}\equiv\left|\dfrac{\left(\Gamma_{12}^s\right)_{SM}}{\left(M_{12}^s\right)_{SM}}
\right|\sin\phi_s=(5.0\pm1.1)\times10^{-3}\,\sin\phi_s\,,
\end{aligned}
\eeq
with
\beq
\begin{aligned}
&\phi_d\equiv\arg\Big(-\left(M_{12}^d\right)_{SM}/\left(\Gamma_{12}^d\right)_{SM}\Big)=
      -4.3^\circ\pm 1.4^\circ\,,\\
&\phi_s\equiv\arg\Big(-\left(M_{12}^s\right)_{SM}/\left(\Gamma_{12}^s\right)_{SM}\Big)=
       0.22^\circ\pm 0.06^\circ\,.
\end{aligned}
\eeq
In order to provide the expression for NP contributions to the semileptonic CP-asymmetry, it is useful 
to adopt a notation for $\Gamma_{12}^q$ similar to that introduced for $M_{12}^q$:
\beq
\Gamma_{12}^q=(\Gamma_{12}^q)_\text{SM}\,\tilde{\mathcal{C}}_{B_q}\,\qquad\text{with}\qquad
\tilde{\mathcal{C}}_{B_q}=1+2\,a_W\,y_b^2\,,
\label{NotationGamma12withNP}
\eeq
in the approximation used here. With such a notation, it follows that 
\beq
a^q_{sl}=\left|\dfrac{\left(\Gamma_{12}^q\right)_{SM}}{\left(M_{12}^q\right)_{SM}}
\right|\dfrac{\tilde{\mathcal{C}}_{B_q}}{\mathcal{C}_{B_q}}\sin\left(\phi_q+2\varphi_{B_q}\right)\,,
\eeq
with $\mathcal{C}_{B_q}$ given in Eq.~(\ref{CBds}).

\section{Phenomenological Analysis}


Given the small corrections to the unitarity of the CKM matrix and to the angles of the unitarity 
triangle described in Eq.~(\ref{angles}), and in particular the tiny and subdominant corrections 
to $\gamma$, it is reasonable to adopt the Wolfenstein parametrization to describe the CKM matrix, 
where the  parameters are fixed considering the value of $V_{us}$, $V_{cb}$, $\gamma$ and $|V_{ub}|$, 
which are related to tree-level processes and therefore hardly affected by NP contributions.

\subsection{The input parameters and the SM analysis}
\label{sec:Input}

\begin{table}[b!]
\renewcommand{\arraystretch}{1}\setlength{\arraycolsep}{1pt}
\center{\begin{tabular}{|l|l|}
\hline
$G_F = 1.16637(1)\times 10^{-5}\GeV^{-2}$\hfill\cite{Nakamura:2010zzi} 	& $m_{B_d}= 5279.5(3)\MeV$\hfill\cite{Nakamura:2010zzi}\\
$M_W = 80.399(23) \GeV$\hfill\cite{Nakamura:2010zzi}  			& $m_{B_s} = 5366.3(6)\MeV$\hfill\cite{Nakamura:2010zzi}\\
$s^2_W\equiv\sin^2\theta_W = 0.23116(13)$\hfill\cite{Nakamura:2010zzi} 	& $F_{B_d} = 205(12)\MeV$\hfill\cite{Laiho:2009eu}\\
$\alpha(M_Z) = 1/127.9$\hfill\cite{Nakamura:2010zzi}				& $F_{B_s} = 250(12)\MeV$\hfill\cite{Laiho:2009eu}\\
$\alpha_s(M_Z)= 0.1184(7) $\hfill\cite{Nakamura:2010zzi}			& $\hat B_{B_d} = 1.26(11)$\hfill\cite{Laiho:2009eu}\\\cline{1-1}
$m_u(2\GeV)=1.7\div3.1\MeV $ 	\hfill\cite{Nakamura:2010zzi}			& $\hat B_{B_s} = 1.33(6)$\hfill\cite{Laiho:2009eu}\\
$m_d(2\GeV)=4.1\div5.7\MeV$	\hfill\cite{Nakamura:2010zzi}			& $F_{B_d} \sqrt{\hat B_{B_d}} = 233(14)\MeV$\hfill\cite{Laiho:2009eu}\\
$m_s(2\GeV)=100^{+30}_{-20} \MeV$	\hfill\cite{Nakamura:2010zzi}		& $F_{B_s} \sqrt{\hat B_{B_s}} = 288(15)\MeV$\hfill\cite{Laiho:2009eu}\\
$m_c(m_c) = (1.279\pm 0.013) \GeV$ \hfill\cite{Chetyrkin:2009fv}		& $\xi = 1.237(32)$\hfill\cite{Laiho:2009eu}\\
$m_b(m_b)=4.19^{+0.18}_{-0.06}\GeV$\hfill\cite{Nakamura:2010zzi} 		& $\eta_B=0.55(1)$\hfill\cite{Buras:1990fn,Urban:1997gw}\\
$M_t=172.9\pm0.6\pm0.9 \GeV$\hfill\cite{Nakamura:2010zzi} 			& $\Delta M_d = 0.507(4) \,\text{ps}^{-1}$\hfill\cite{Nakamura:2010zzi}\\\cline{1-1}
$m_K= 497.614(24)\MeV$	\hfill\cite{Nakamura:2010zzi}				& $\Delta M_s = 17.77(12) \,\text{ps}^{-1}$\hfill\cite{Nakamura:2010zzi}\\	
$F_K = 156.0(11)\MeV$\hfill\cite{Laiho:2009eu}				& $\sin(2\beta)_{b\to c\bar c s}= 0.673(23)$\hfill\cite{Nakamura:2010zzi}\\
$\hat B_K= 0.737(20)$\hfill\cite{Laiho:2009eu}				& $\phi_s^{\psi\phi}= 0.55^{+0.38}_{-0.36}$\hfill\cite{Giurgiu:2010is,Abazov:2011ry}\\
$\kappa_\epsilon=0.923(6)$\hfill\cite{Blum:2011ng}				& $\phi_s^{\psi\phi}= 0.03\pm0.16\pm0.07 $\hfill\cite{Koppenburg:PC}\\	
$\varphi_\epsilon=(43.51\pm0.05)^\circ$\hfill\cite{Buras:2008nn}		& $R_{\Delta M_B}=(2.85\pm0.03)\times 10^{-2}$\hfill\cite{Nakamura:2010zzi}\\\cline{2-2}
$\eta_1=1.87(76)$\hfill\cite{Brod:2011ty}					& $A^b_{sl}=(-0.787\pm0.172\pm0.093)\times10^{-2}$\hfill\cite{Abazov:2011yk}	\\\cline{2-2}		
$\eta_2=0.5765(65)$\hfill\cite{Buras:1990fn}					& $|V_{us}|=0.2252(9)$\hfill\cite{Nakamura:2010zzi}\\
$\eta_3= 0.496(47)$\hfill\cite{Brod:2010mj}					& $|V_{cb}|=(40.6\pm1.3)\times 10^{-3}$\hfill\cite{Nakamura:2010zzi}\\ 
$\Delta M_K= 0.5292(9)\times 10^{-2} \,\text{ps}^{-1}$\hfill\cite{Nakamura:2010zzi}	& $|V^\text{incl.}_{ub}|=(4.27\pm0.38)\times10^{-3}$\hfill\cite{Nakamura:2010zzi}\\
$|\eps_K|= 2.228(11)\times 10^{-3}$\hfill\cite{Nakamura:2010zzi}		& $|V^\text{excl.}_{ub}|=(3.38\pm0.36)\times10^{-3}$\hfill\cite{Nakamura:2010zzi}\\\cline{1-1}
$\tau_{B^\pm}=(1641\pm8)\times10^{-3}\ps$\hfill\cite{Nakamura:2010zzi}	& $|V^\text{comb.}_{ub}|=(3.89\pm0.44)\times10^{-3}$\hfill\cite{Nakamura:2010zzi}\\																				
$BR(B^+\to\tau^+\nu)=(1.65\pm0.34)\times10^{-4}$\hfill\cite{Nakamura:2010zzi}& $\gamma=(73^{+22}_{-25})^\circ$\hfill\cite{Nakamura:2010zzi}\\\hline
\end{tabular}} 
\caption{\it Values of the experimental quantities used as input parameters. Notice that $m_i(m_i)$ are 
the masses $m_i$ at the scale $m_i$ in the $\ov{MS}$ scheme while $M_t$ is the top-quark pole mass. 
\label{tab:input}}
\renewcommand{\arraystretch}{1.0}
\end{table}
The physical parameters considered in our analysis and their present experimental values are summarized 
in Table~\ref{tab:input}. First of all, notice that $V_{us}$ and $V_{cb}$ appear to be relatively well 
measured, compared to $V_{ub}$ which has an error still of the order of 10\%. Moreover there is a latent 
tension between the exclusive and the inclusive experimental determinations of $|V_{ub}|$, which translates 
into the well known $\vep_K-S_{\psi K_S}$ and $BR(B^+\to\tau^+\nu)$ anomalies. Finally the angle $\gamma$ 
of the $B_d$ unitarity triangle, despite of being a tree-level SM processes, still suffers from a large 
uncertainty. {\em Rebus sic stantibus}, two different scenarios may be depicted, either the inclusive or 
the exclusive determination of $|V_{ub}|$ is preferred:
\begin{itemize}
\item[i)]
Using the {\it exclusive determination of $|V_{ub}|$}, $S_{\Psi K_S}$ is predicted to be very close to 
the experimental determination of $\sin(2\beta)_{b\to c\bar c s}$, while $\vep_K\approx1.8\times10^{-3}$ 
is clearly below the measured value. Furthermore, for such a value of $|V_{ub}|$ the $BR(B^+\to\tau^+\nu)$ 
is predicted to be $0.85\times10^{-4}$, smaller than the central experimental value shown in 
Table~\ref{tab:input} by more than $2\sigma$. If NP is advocated in order to solve (or at least to soften) 
these anomalies it should enhance the value of $\vep_K$ and $BR(B^+\to\tau^+\nu)$, while having negligible 
impact on $S_{\psi K_S}$.
\item[ii)]
Using {\it the inclusive determination of $|V_{ub}|$}, the SM prediction for $\vep_K$ is closer to its 
experimental determination and $BR(B^+\to\tau^+\nu)\simeq1.35\times10^{-4}$ agrees with the measured 
value within the $1\sigma$ level. However, $S_{\psi K_S}\approx0.8$ is above the measured value. If NP 
is advocated in order to solve (or at least to soften) this anomaly , it should deplete $S_{\psi K_S}$, 
while leaving basically unchanged $\vep_K$ and $BR(B^+\to\tau^+\nu)$.
\end{itemize}

In the previous section, we have shown that NP contributions to $S_{\psi K_S}$, encoded in $\varphi_{B_d}$, 
are negligible, being suppressed by $y_b^2$, while NP contributions to $\vep_K$ are sizable. As a result, 
our framework seems to fit better inside scenario i) and, consequently, in the following we will assume 
the exclusive determination of $|V_{ub}|$ to be the ``correct'' one.  

\begin{figure}[t!]
 \centering
 \includegraphics[width=11cm]{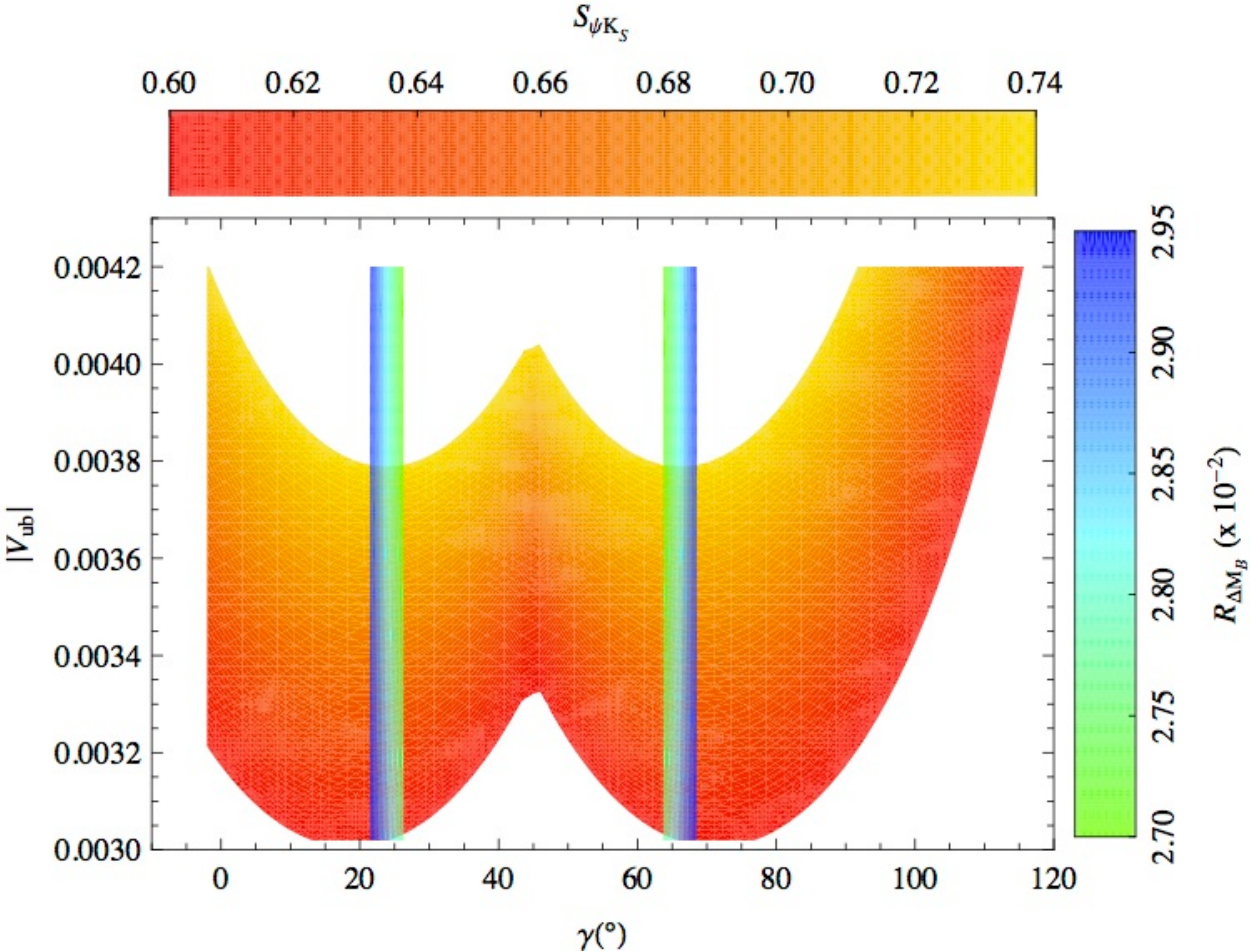}
 \caption{\it $|V_{ub}|-\gamma$ parameter space for which $S_{\psi K_S}$ (orange-red tone) and 
   $R_{\Delta M_B}$ (blue-green tone) are inside the corresponding $3\sigma$ error ranges.}
 \label{fig:VubGamma}
\end{figure}

Assuming the SM, it is possible to consider the constraints on the $|V_{ub}|-\gamma$ parameter space, 
deriving from the independent measurements of $R_{\Delta M_B}$ and $S_{\psi K_S}$. The present situation 
is depicted in Fig.~\ref{fig:VubGamma}, for $S_{\psi K_S}$ (orange-red tone) and $R_{\Delta M_B}$ 
(green-blue tone) inside their corresponding $3\sigma$ error ranges (see Table~\ref{tab:input}). The 
figure shows that $R_{\Delta M_B}$ strongly reduces the allowed parameter space for $\gamma$ to two 
intervals: the first one is $\sim[21.5^\circ,\,26.3^\circ]$, which is around the experimental lower 
$2\sigma$ value; the second one is $\sim[63.7^\circ,\,68.5^\circ]$, very close to the central value. 
In the next years, LHCb will lower the uncertainty on $\gamma$ and hopefully one of these two regions 
will be excluded. The figure also illustrates that $S_{\psi K_S}$ constrains $|V_{ub}|$ only to relatively 
small values, $\sim[3.0\times10^{-3},\,3.8\times10^{-3}]$, a pattern already mentioned above when 
discussing the exclusive analysis of $|V_{ub}|$. In the following we will refer to allowed $|V_{ub}|-\gamma$ 
parameter space, once included the combined bounds from $R_{\Delta M_B}$ and $S_{\psi K_S}$, as the 
``reduced'' $|V_{ub}|-\gamma$ parameter space.

\begin{figure}[t!]
 \centering
\subfigure[$\vep_K$]{
 \includegraphics[width=7.7cm]{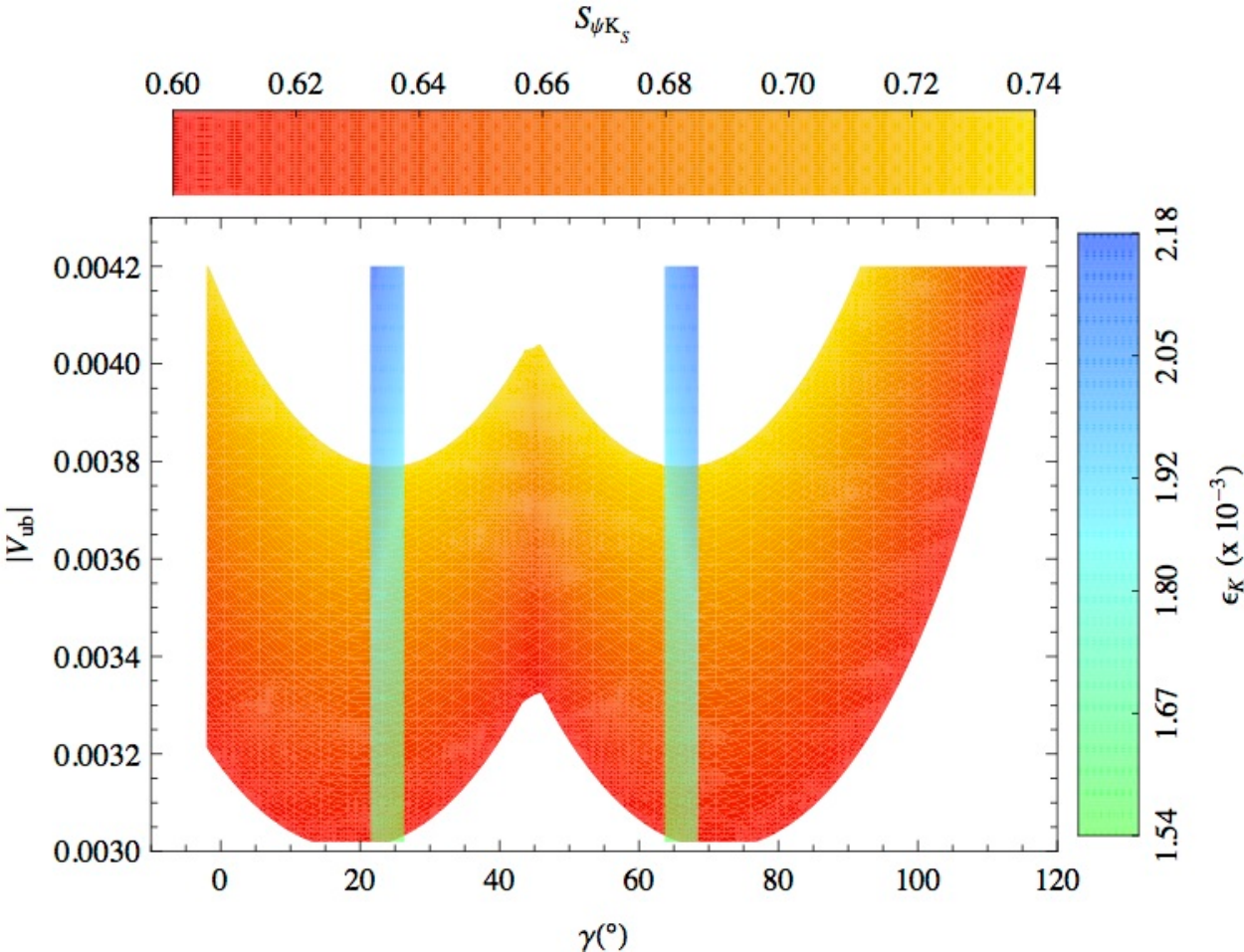}
}
 \subfigure[$\Delta M_{B_d}$]{
 \includegraphics[width=7.7cm]{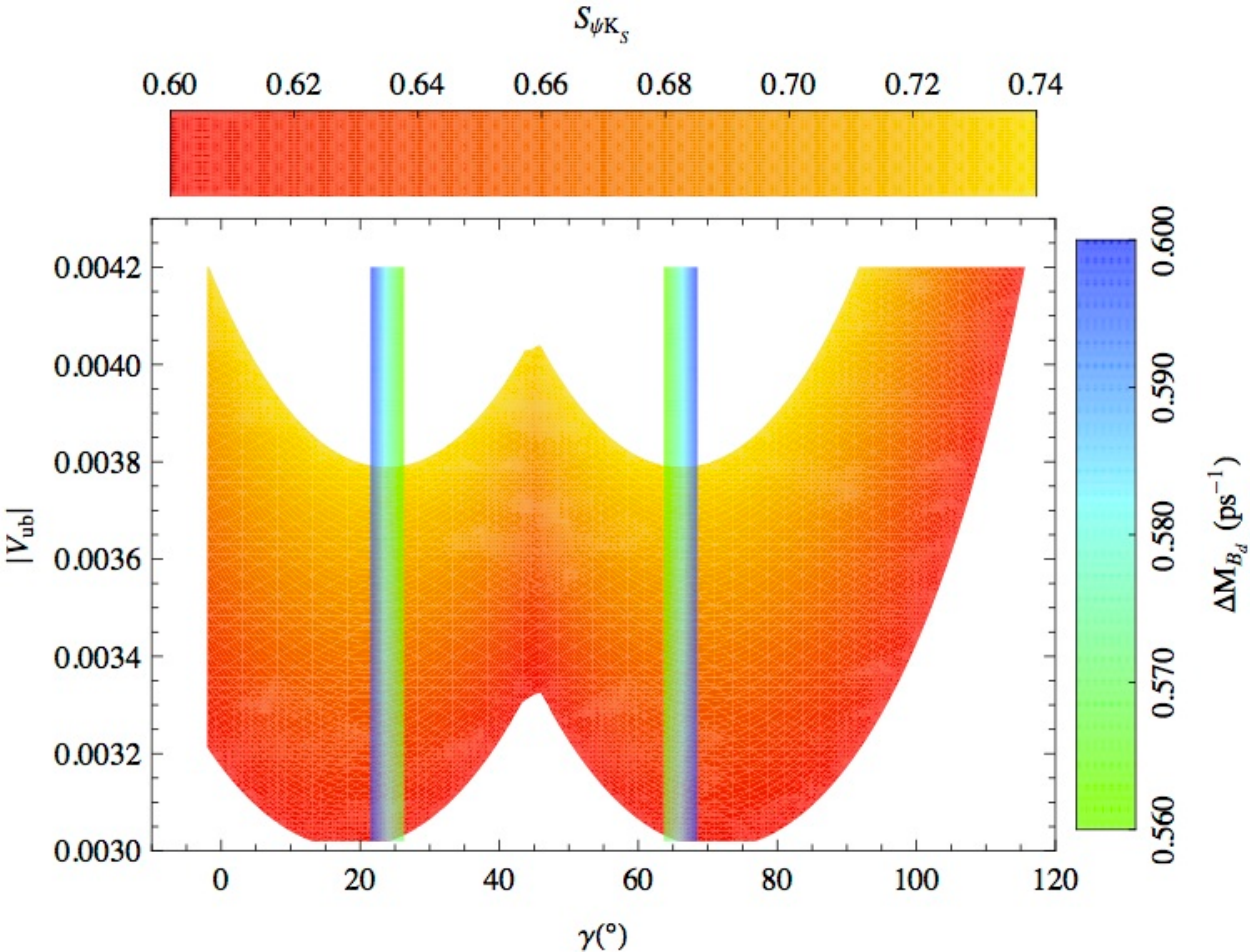}
}
\subfigure[$\Delta M_{B_s}$]{
 \includegraphics[width=7.7cm]{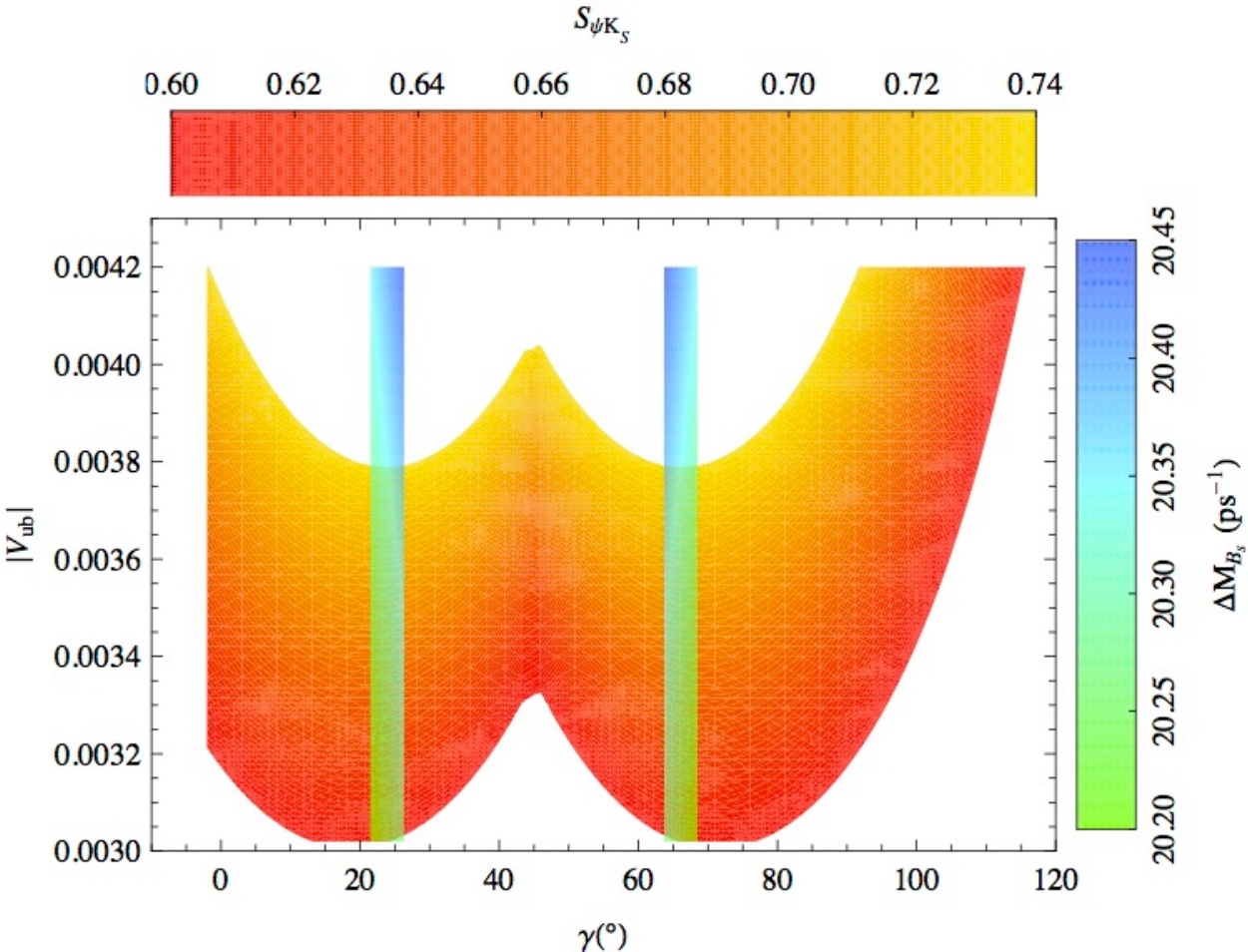}
}
  \subfigure[$R_{BR/\Delta M}$]{
 \includegraphics[width=7.7cm]{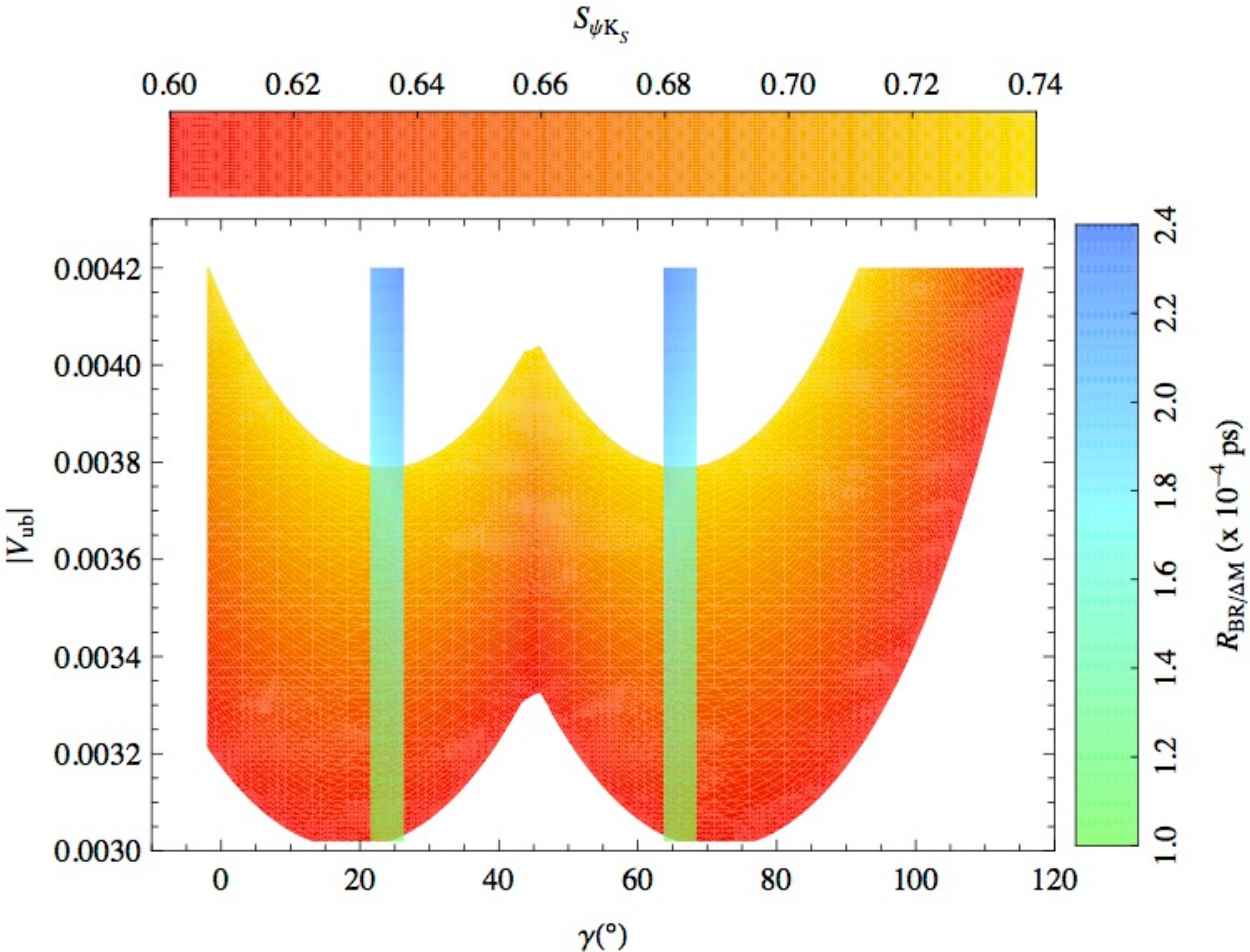}
}
\caption{\it SM predictions for $\vep_K$, $\Delta M_{B_{d,s}}$ and $R_{BR/\Delta M}$ in the reduced 
             $|V_{ub}|-\gamma$ parameter space.}
\label{fig:VubGammaObservables}
\end{figure}

It Fig.~\ref{fig:VubGammaObservables} the variation of the SM predictions for $\vep_K$, $\Delta M_{B_{d,s}}$ 
and $R_{BR/\Delta M}$ in the ``reduced'' $|V_{ub}|-\gamma$ parameter space is studied. In each of the 
sub-plots of Fig.~\ref{fig:VubGammaObservables} the variation of the SM central values predictions is 
shown, with the colors corresponding to the legend on the right side of each subplot. While the SM 
prediction for $\vep_K$, Fig.~\ref{fig:VubGammaObservables} (a), turns out to be always smaller than 
its experimental determination, the SM predictions for the mass differences $\Delta M_{B_d}$, 
Fig.~\ref{fig:VubGammaObservables} (b), and  $\Delta M_{B_d}$, Fig.~\ref{fig:VubGammaObservables} (c)
are above the corresponding data. However, given the large theoretical uncertainties on these quantities, 
this tendency should not be considered more than just a ``slight indication''. The ratio $R_{BR/\Delta M}
\equiv BR(B^+\to\tau^+\nu)/\Delta M_{B_d}$ can be useful in order to reduce most of the theoretical 
uncertainties on $\Delta M_{B_d}$. Its SM prediction can be obtained from Eq.~(\ref{R_BRoverDM}) setting 
$a_W=a_{CP}=0$. Fig.~\ref{fig:VubGammaObservables} (d) depicts the variation of the SM prediction in the 
``reduced'' $|V_{ub}|-\gamma$ parameter space. Even if always inside the $3\sigma$ error range, it appears  
evident the SM preference for a small $R_{BR/\Delta M}$ ratio, close to the lower experimental limit:
\beq
\left(R_{BR/\Delta M}\right)_{exp}=(3.25\pm0.67)\times 10^{-4}\ps\,.
\eeq

These particular patterns of the SM predictions (always either above or below or inside the $3\sigma$ 
error ranges) are going to be relevant when discussing NP effects, because it is a common feature of 
all the points in the ``reduced'' $|V_{ub}|-\gamma$ parameter space. 

In order to illustrate the features of the MFV scenario with a strong interacting Higgs sector, the 
numerical analysis of the following sections will be presented choosing as reference point, 
$(|V_{ub}|,\,\gamma)=(3.5\times 10^{-3},\,66^\circ)$, corresponding to $S_{\psi K_S}\simeq0.692$ and 
$R_{\Delta M_B}\simeq2.83\times10^{-2}$. For this point, $\gamma$, $S_{\psi K_S}$, $R_{\Delta M_B}$ 
and $|V_{ub}|$ are inside their own 1$\sigma$ error determinations\footnote{We are considering the 
exclusive determination of $|V_{ub}|$ as commented in the previous section.}.

\boldmath
\subsection{FCNC Constraints on $a_{CP}$, $a_{W}$ and $a_Z^d$}
\label{sec:PhenoAnalysis}
\unboldmath

The analysis presented here mainly relies on two observables: $\vep_K$ and $R_{BR/\Delta M}$. Indeed, as seen 
in the previous section, $S_{\psi K_S}$ and $R_{\Delta M_B}$ turn out to be only slightly affected by NP and 
very close to their SM predictions. On the other hand, $\Delta M_K$ and $\Delta M_{B_{d,s}}$ are affected 
by large theoretical errors related to the long distance contributions. Therefore the only requirement on 
the neutral meson mass differences will be that $(\Delta M_K)_{exp}$ is reproduced within $\pm40\%$. Finally, 
the discussion on $S_{\psi \phi}$ and $A^b_{sl}$ is presented separately. 

As none of the observables considered here get contributions from $a_Z^u$, the analysis will be restricted 
to the remaining three parameters $a_{CP}$, $a_{W}$ and $a_Z^d$. The analytic expressions for the NP 
contribution to the considered observables are deferred to App.~\ref{App:PhemSection}.

Before entering into the detailed determination of the bounds on $a_{CP}$, $a_{W}$ and $a_Z^d$ from FCNC 
constraints, it is pertinent to point out that further constraints on the $d_{\chi=4}$ operator coefficients 
of the non-linear expansion may result from data on flavour-conserving transitions, such as for instance 
from the constraints on the oblique parameters $S$, $T$, $U$ \cite{Peskin:1990zt,Peskin:1991sw}. A rough 
order-of-magnitude estimate of the NP effects under consideration is given by the correction to the 
$T$ parameter,
\be
T^{NP} - T^{SM} \approx  a_X y_t^2\, T^{SM} 
\ee
where $T^{NP}$ is the total SM contribution due to the top and the NP loop, with $a_X$ denoting a generic 
NP coefficient of the $d_{\chi=4}$ Lagrangian Eq.~(\ref{DevSM}), and  
\be
T^{SM} \approx \frac{3 G_F M_t^2}{8\pi^2\alpha \sqrt{2}}\,.
\ee
The NP effects may be traded by a variation on the value of the quark top mass in the SM contribution. 
A change of the top mass of order of $5\%$ produces a change of the $T$ parameter of $0.1$, well inside 
the 2$\sigma$ LEP result (see Fig. E.2  in Ref.~\cite{:2005ema}). This translates into a bound of order 
$5-10\%$ on the couplings $a_X$. A thorough and detailed analysis of the one-loop impact on EW precision 
measurements is beyond the scope of the present paper and will be presented elsewhere~\cite{USworkinprogress}.

\mathversion{bold}
\subsubsection{$\vep_K$ and $R_{BR/\Delta M}$ }
\mathversion{normal}
For the chose reference point in the $(|V_{ub}|,\,\gamma)$ parameter space, the central values for the SM 
predictions are:
\beq
\vep_K=1.8\times 10^{-3}\,,\qquad\qquad
R_{BR/\Delta M}=1.6\times10^{-4}\ps\,.
\label{SMpredictionsVub}
\eeq
As anticipated before in Fig.~\ref{fig:VubGammaObservables}, the SM prediction for $\vep_K$ is below 
the experimental value, while that for $R_{BR/\Delta M}$ is already inside but close to the 3$\sigma$ 
error bound. The main question is if the MFV scenario with strong Higgs dynamics discussed here is able 
to accommodate contemporaneously both  observables: in other words, whether $\vep_K$ can be enhances 
by NP while not spoiling $R_{BR/\Delta M}$.

The NP corrections to $\vep_K$ and $R_{BR/\Delta M}$ turn out to be inversely correlated in our scenario: 
when $\vep_K$ increases  $R_{BR/\Delta M}$ decreases, and vice-versa. More quantitatively, considering only 
one non-vanishing NP parameter at time we find the following behavior:
\beq
\begin{aligned}
&a_{CP}=\pm0.1 &&\longrightarrow\quad && \delta\vep_K\approx1.1\% \,,\quad&& \delta R_{BR/\Delta M}\approx-1.4\%\,,\\
&a_W\,\,=0.1(-0.1) &&\longrightarrow\quad && \delta\vep_K\approx+26\%(-19\%)\,, \quad&& \delta R_{BR/\Delta M}\approx-25\%(+30\%)\,,\\
&a_Z^d\,\,\,=\pm0.1 &&\longrightarrow\quad && \delta\vep_K\approx124\% \,,\quad&& \delta R_{BR/\Delta M}\approx-62\%\,.
\end{aligned}
\label{DependenceNP}
\eeq
Notice that $\vep_K$ can be suppressed only for negative values of $a_W$, while it is always enhanced 
for positive ones; exactly the opposite happens for $R_{BR/\Delta M}$. Furthermore, $\vep_K$ and 
$R_{BR/\Delta M}$ are {\it a priori} very sensitive to the tree level FCNC $Z$-exchange contributions, 
parametrized by $a_Z^d$. However, this parameter is also strongly constrained by $\Delta F=1$ observables, 
as shown in sec. \ref{sec:DeltaF1}. Finally, we have also checked that this pattern is do not strongly 
rely on the particular point chosen in the $|V_{ub}|-\gamma$ parameter space.
\begin{figure}[h!]
 \centering
 \subfigure[Correlation plot between $\vep_K$ and $R_{BR/\Delta M}$. $a_W=a_Z^d=0$, $a_{CP}\in{[}-1,1{]}$.]{
\includegraphics[width=7cm]{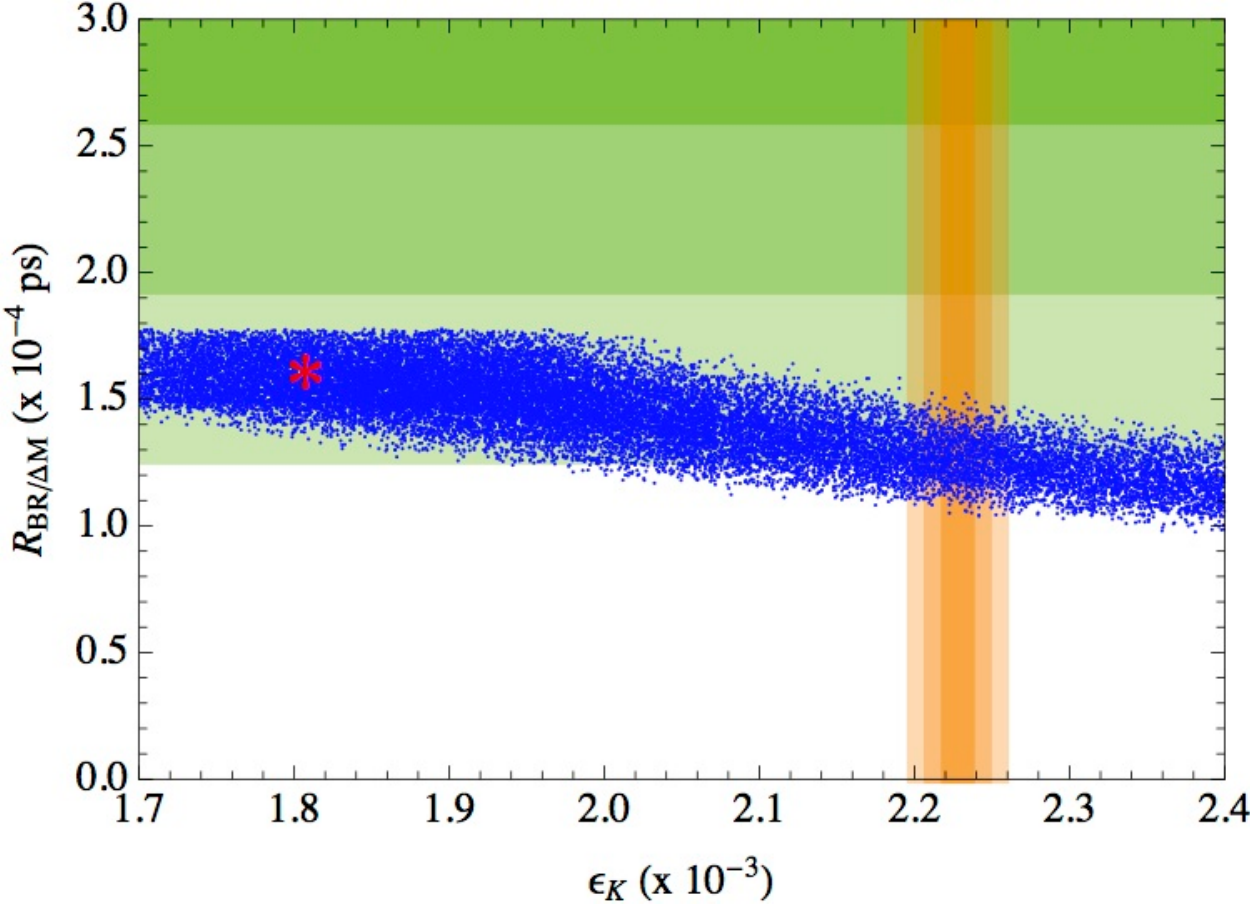}}
 \subfigure[$a_{CP}$ parameter space in terms of $\vep_K$ for $R_{BR/\Delta M}$ insider its $3\sigma$ 
  error range.]{
\includegraphics[width=7cm]{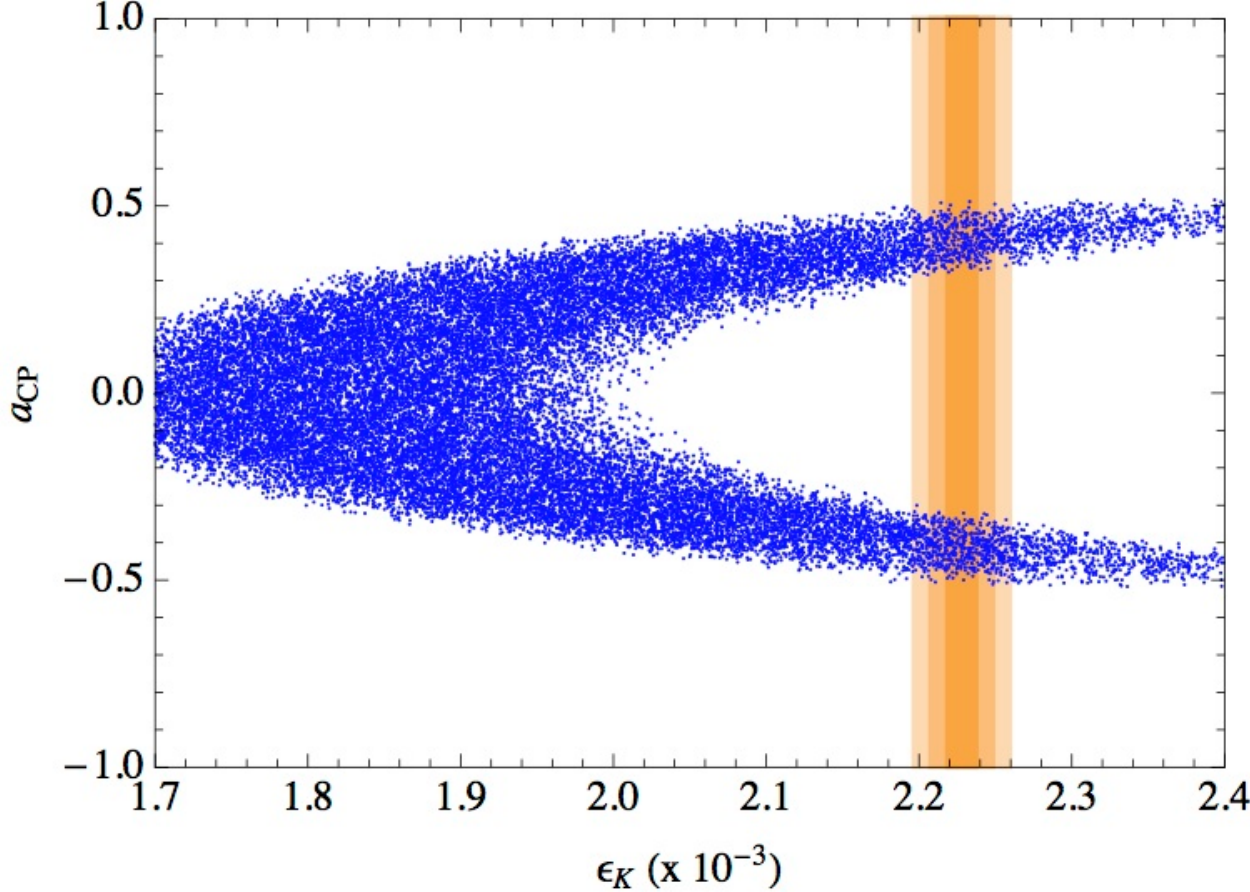}}
 \subfigure[Correlation plot between $\vep_K$ and $R_{BR/\Delta M}$. $a_W\in{[}-1,1{]}$, 
$a_Z^d\in{[}-0.1,0.1{]}$ and $a_{CP}=0$.]{
\includegraphics[width=7cm]{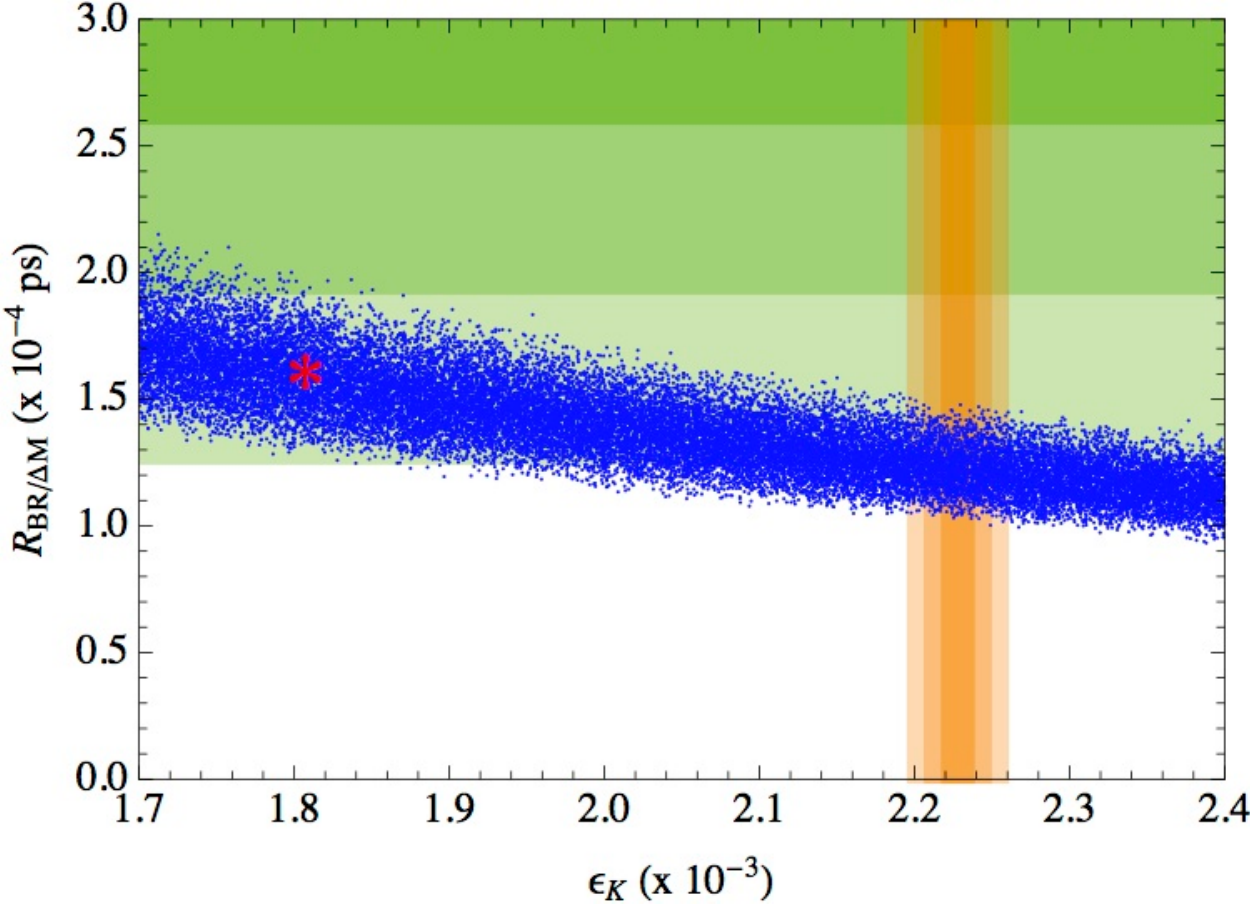}}
 \subfigure[$a_W-a_Z^d$ parameter space for the observables inside their $3\sigma$ error ranges.]{
\includegraphics[width=7cm]{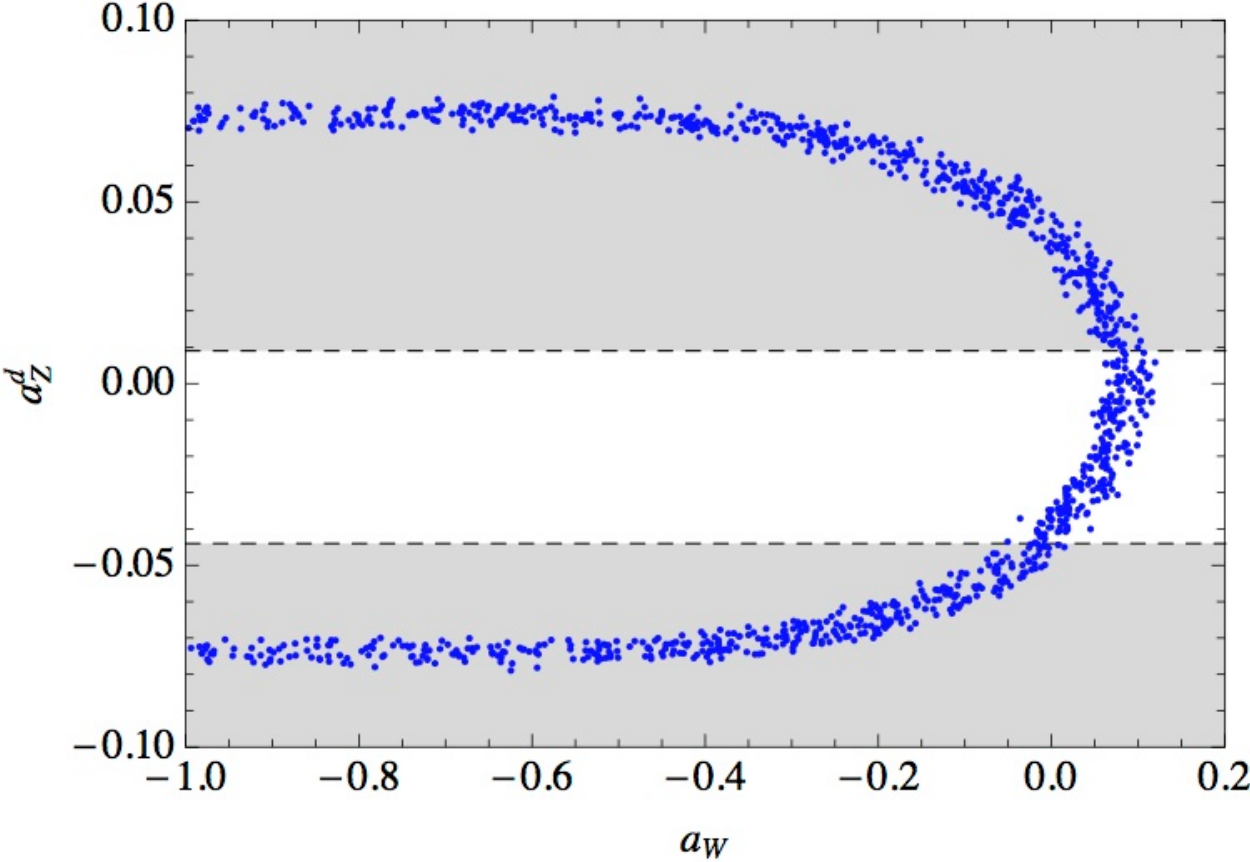}}
 \subfigure[Correlation plot between $\vep_K$ and $R_{BR/\Delta M}$. $a_W,a_{CP}\in{[}-1,1{]}$, 
            $a_Z^d\in {[}-0.1,0.1{]}$]{
\includegraphics[width=7cm]{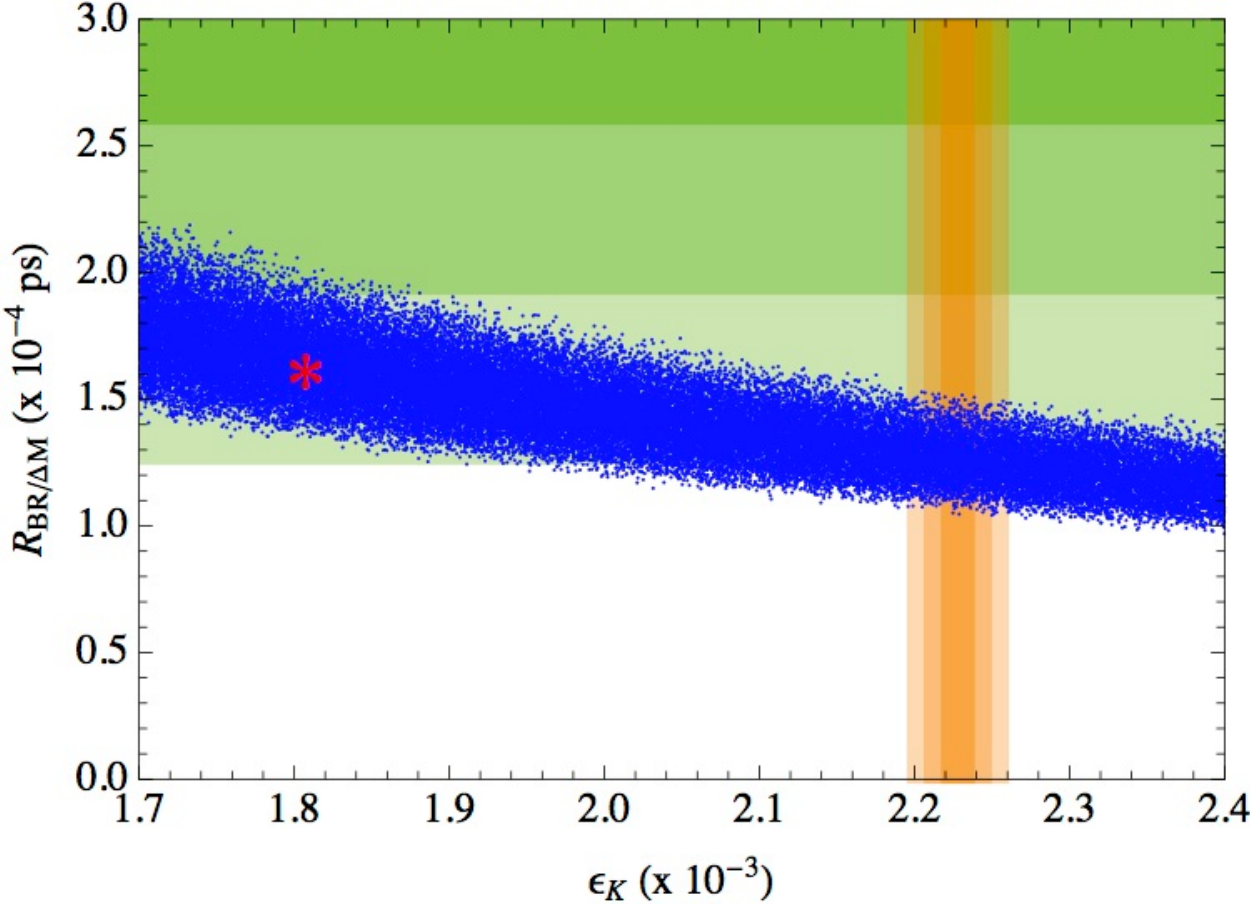}}  
\subfigure[$a_W-a_{CP}$ parameter space for the observables inside their $3\sigma$ error ranges and 
            $a_Z^d\in{[}-0.044,0.009{]}$.]{
\includegraphics[width=7cm]{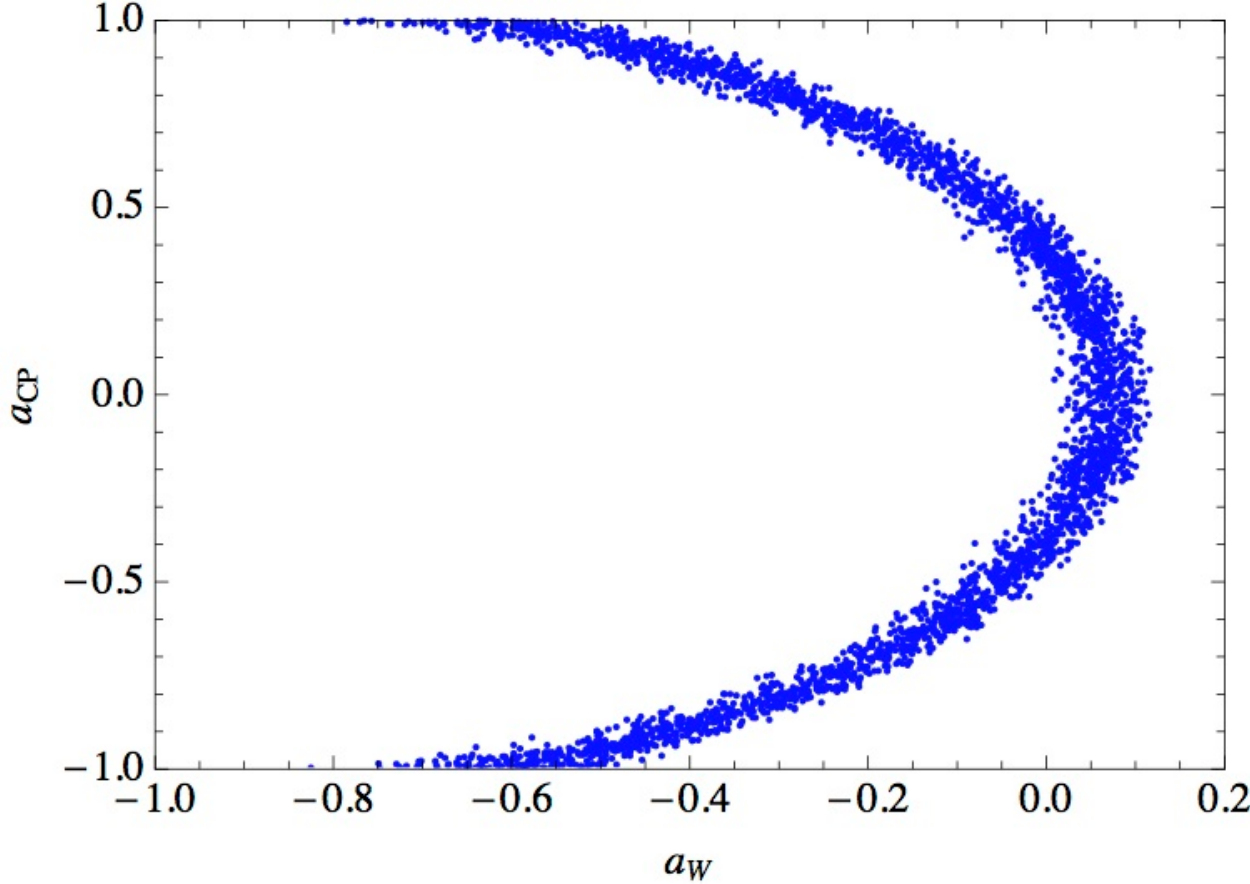}}
\caption{\it Results for the reference point $(|V_{ub}|,\,\gamma)=(3.5\times 10^{-3},\,66^\circ)$. 
  Details in the text.}
\label{fig:EpsilonKRatio}
\end{figure}

In Fig.~\ref{fig:EpsilonKRatio} we show the results considering the NP contributions and the input 
parameters (see Table~\ref{tab:input}). We consider three different cases:
\begin{enumerate}
\item{Only $a_{CP}$ is taken to be non-vanishing, see plots (a) and (b) of Fig.~\ref{fig:EpsilonKRatio}. 
This case shows the impact of the CP odd operator by itself. Plot (a) illustrates the $\vep_K$-
$R_{BR/\Delta M}$ correlation: in orange (green) 
the $1\sigma$, $2\sigma$ and $3\sigma$ (from the darker to the lighter) experimental error ranges for 
$\vep_K$ ($R_{BR/\Delta M}$), in blue the correlation between $\vep_K$ and $R_{BR/\Delta M}$, while 
the red star represents the SM predictions in Eq.~(\ref{SMpredictionsVub}). Plot (b) depicts the $a_{CP}$ 
parameter space in terms of $\vep_K$ for $R_{BR/\Delta M}$ inside its $3\sigma$ error range: only 
$|a_{CP}|$ values in the range $\sim[0.3,\,0.5]$ can accommodate both observables. }
\item{Only $a_W$ and $a_Z^d$ parameters are taken to be non-vanishing, see plots (c) and (d) of 
Fig.~\ref{fig:EpsilonKRatio}. This illustrates the case when only the CP even operators are taken into 
account. This case can be considered as the strict MFV limit~\cite{D'Ambrosio:2002ex} of our scenario: 
the CP-invariance of the Lagrangian is recovered and there are no new CP-violating sources other than the 
CKM one. Plot (c) is the analogous of plot (a) and the same description applies. 
Furthermore, those two plots are very similar and therefore the impact of $a_{CP}$ alone and of $a_W$ 
and $a_Z^d$ together is comparable. Plot (d) shows the $a_W-a_Z^d$ parameter space: the blue points 
refer to $\vep_K$ and $R_{BR/\Delta M}$ inside their $3\sigma$ error ranges; the gray regions represent 
the $\Delta F=1$ bounds on $a_Z^d$ parameter, $a_Z^d\in[-0.044,0.009]$, and as a result a large part 
of the $a_W-a_Z^d$ parameter space is excluded: this puts a strong bound on $a_W$, which can take values 
only inside a narrow range $~[-0.1,\,0.1]$.}
\item{All parameters are considered to be simultaneously non-vanishing, se plots (e) and (f) of 
Fig.~\ref{fig:EpsilonKRatio}. Plot (e) is very similar to plots (a) and (c) and therefore the same 
conclusions apply. Plot (f) shows the $a_W-a_{CP}$ parameter space, in which the blue points refer 
to $\vep_K$ and $R_{BR/\Delta M}$ inside their $3\sigma$ error ranges and $a_Z^d$ is inside the bound 
coming from the $\Delta F=1$ observable analysis. Only a small part of the $a_W-a_{CP}$ parameter space 
survives: it is however interesting to notice that $|a_W|$ and $|a_{CP}|$ can take even large values 
and this is the result of a partial cancellation among the contributions of $a_{CP}$ and $a_W$, as we 
have explicitly verified.}
\end{enumerate}
Summarizing, our MFV scenario with a strong interacting Higgs sector is able to accommodate simultaneously 
$\vep_K$, $R_{BR/\Delta M}$, $S_{\psi K_S}$, and $R_{\Delta M_B}$, only including the theoretical errors. 
The NP parameter space is strongly constrained and only for a small part of it all the observables take 
values inside their own experimental $3\sigma$ errors ranges. Correlations between contributions from 
$a_W$ and $a_{CP}$, even large values for these two parameters are allowed. Conversely, $a_Z^d$ is strongly 
constrained, mainly from $\Delta F=1$ observables. 

For the parameter space region selected by all considered constraints, the non-linear realization of MFV 
is able to solve the $\vep_K-S_{\psi K_S}$ anomaly, but it is unable to remove the SM tension on 
$BR(B^+\to\tau^+\nu)$. However, a better agreement with the data can be found selecting different values 
for  $|V_{ub}|$ and $\gamma$: indeed, slightly larger values for $|V_{ub}|$, while keeping $\epsilon_K$, 
$R_{\Delta M_B}$ and $S_{\psi K_S}$ in agreement with data at $1\sigma$, would enhance $BR(B^+\to\tau^+\nu)$ 
towards its experimental central value.

\mathversion{bold}
\subsubsection{$S_{\psi \phi}$ and $A^b_{sl}$}
\mathversion{normal}

The mixing-induced CP asymmetry in the $B^0\to\psi\phi$ decay has been measured by D0 , which recently 
updated its analysis with $8\fb^{-1}$ of data~\cite{Giurgiu:2010is,Abazov:2011ry}, and by LHCb, which 
has presented its preliminary measurement~\cite{Koppenburg:PC}. The numerical results have been included 
in Table~\ref{tab:input} and translate into the $S_{\psi\phi}$ asymmetry values:
\beq
D0:\, S_{\psi\phi}=0.52^{+0.23}_{-0.33} \quad \longleftrightarrow \quad 
LHCb:\,S_{\psi\phi}=0.13\pm0.18\pm0.07\,,
\eeq
which are in agreement, even if the two central values are relatively different. In particular, the LHCb 
result is closer to the SM value, that for our $(|V_{ub}|,\,\gamma)$ reference point  turns out to have 
the central value
\beq
S_{\psi\phi}=0.036\,.
\label{SpsiphiSMPrediction}
\eeq 
Alike to the $S_{\psi K_S}$ asymmetry previously analyzed, in our scenario the NP contributions to 
$S_{\psi\phi}$ are  expected to be suppressed by $y_b^2$-dependent terms, as shown in Eq.~(\ref{NPphaseB}), 
and  indeed our numerical analysis predicts only negligible deviations from the SM value.

The situation is different when the $B$ semileptonic CP-asymmetry $A^b_{sl}$ is considered. The central 
value of the SM prediction, for our $(|V_{ub}|,\,\gamma)$ reference point, is given by: 
\beq
A^b_{sl}=-2.3\times10^{-4}\qquad\qquad 
\left(a^d_{sl}=-4.0\times10^{-4}\,,\qquad a^s_{sl}=1.9\times10^{-5}\right)\,,
\label{AbSLSMPrediction}
\eeq
to be compared with the experimental determination by D0 \cite{Abazov:2011yk}, 
$A^b_{sl}=(-78.7 \pm 17.2\pm9.3)\times10^{-4}$, which deviates from the SM prediction by more than 
$3\sigma$. The question now is whether the scenario explored here may provide contributions to $A^b_{sl}$ 
of the correct sign and sufficiently large to soften the tension with respect to the experimental data.

It turns out that indeed $A^b_{sl}$ could receive sizable NP contributions. For instance, for our reference 
$(|V_{ub}|,\,\gamma)$ values, it results: 
\beq
\begin{aligned}
&a_{CP}=0.1(-0.1) &&\longrightarrow\qquad && \delta A^b_{sl}\approx1.1\%(1.6\%) \\
&a_W\,\,=0.1(-0.1) &&\longrightarrow\qquad && \delta A^b_{sl}\approx33\%(-23\%) \\
&a_Z^d\,\,\,=\pm0.1 &&\longrightarrow\qquad && \delta A^b_{sl}\approx160\%\,.
\end{aligned}
\eeq
 This NP sensitivity of $A^b_{sl}$ is not a common feature of NP models as it has been discussed in 
Ref.~\cite{Lenz:2011zz} and therefore it is worth to further investigate it numerically.
\begin{figure}[h!]
 \centering
  \subfigure[$a_W=a_Z^d=0$, $a_{CP}\in{[}-1,1{]}$.]{
\includegraphics[width=7.5cm]{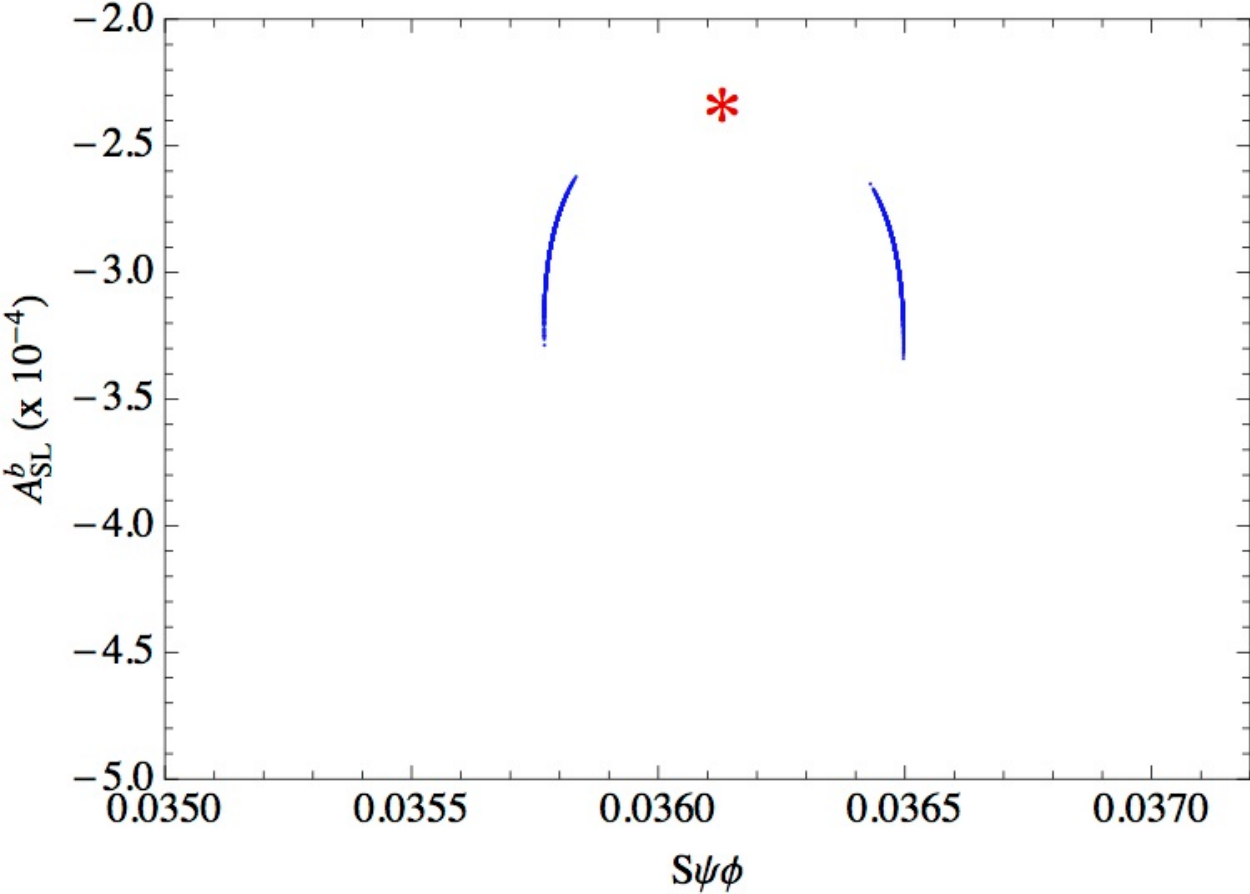}}
  \subfigure[$a_W\in{[}-1,1{]}$, $a_{CP}=0$, $a_Z^d\in{[}-0.044,0.009{]}$.]{
\includegraphics[width=7.5cm]{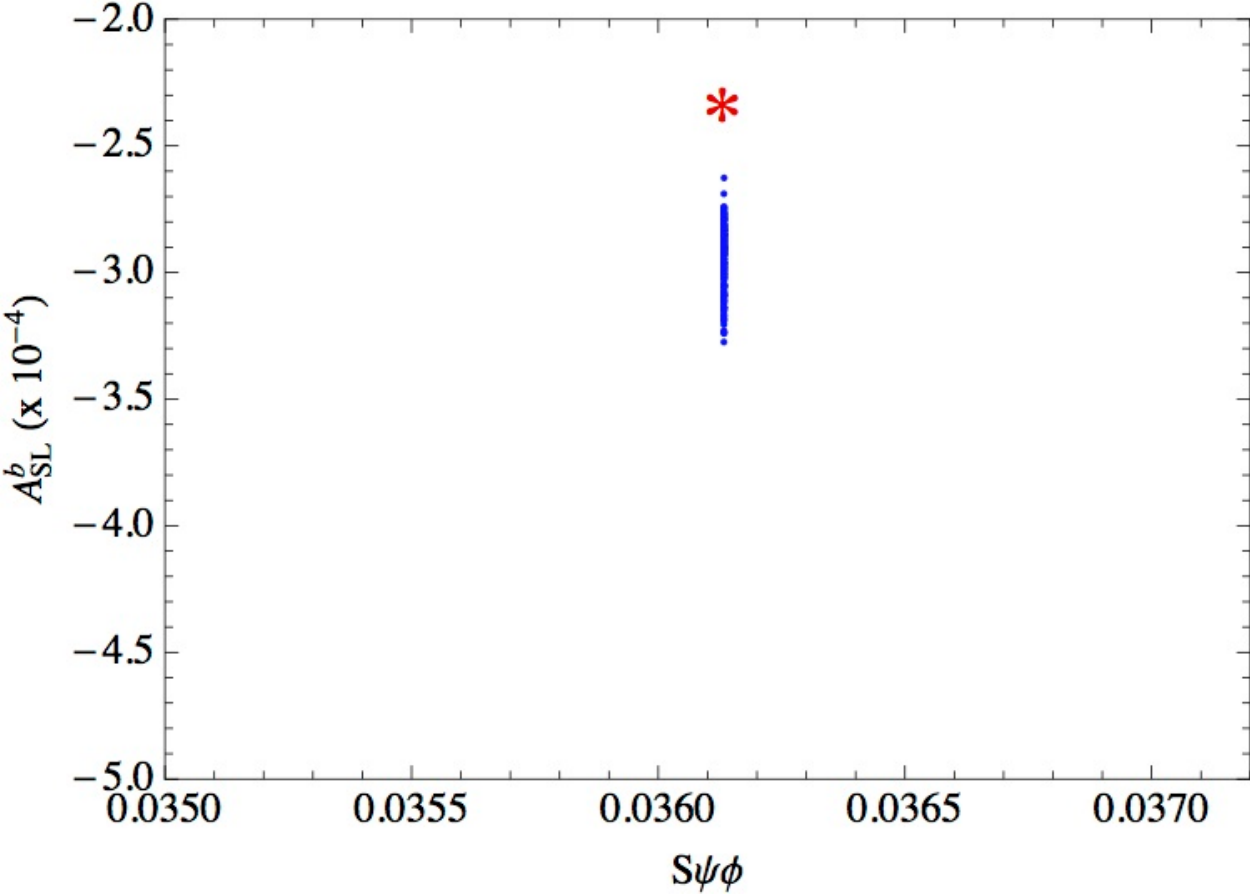}}
  \subfigure[$a_W,a_{CP}\in{[}-1,1{]}$, $a_Z^d\in{[}-0.044,0.009{]}$.]{
\includegraphics[width=7.5cm]{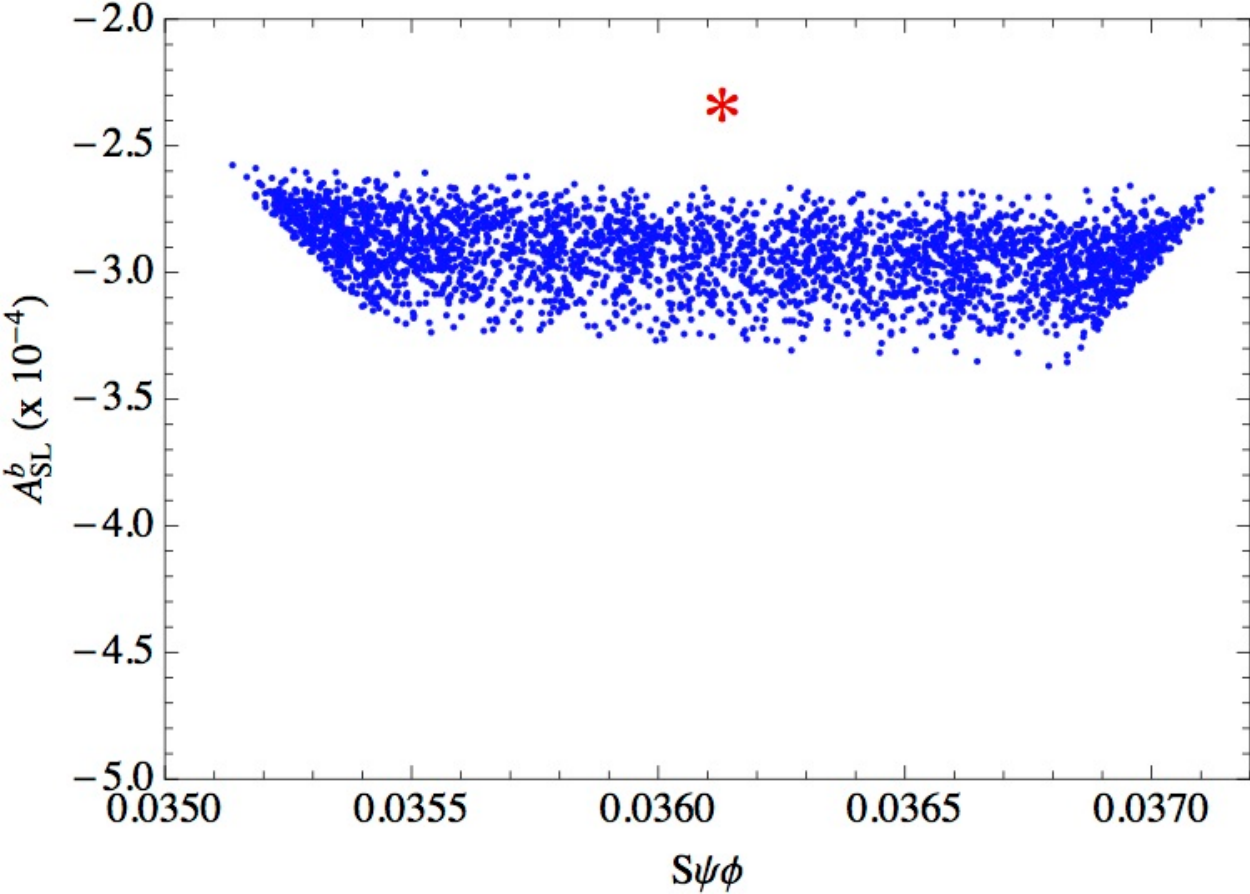}}
\caption{\it Correlation between $A^b_{sl}$ and $S_{\psi\phi}$.  For all points, $\vep_K$ and 
         $R_{BR/\Delta M}$ are inside their $3\sigma$ error ranges. See the text for a detailed description.}
\label{fig:AbSLcorrelation}
\end{figure}

Fig.~\ref{fig:AbSLcorrelation} illustrates the correlation between $A^b_{sl}$ and $S_{\psi\phi}$, always 
for the reference $(|V_{ub}|,\,\gamma)$ value, and constraining $\vep_K$ and $R_{BR/\Delta M}$ to remain 
inside their $3\sigma$ error ranges and $a_Z^d$ to satisfy the $\Delta F=1$ bounds. Again it is worth to 
explore the three different cases:
\begin{enumerate} 
\item{Only $a_{CP}$ is considered to be non-vanishing, see plot (a) of Fig.~\ref{fig:AbSLcorrelation}, 
spanning  the range $[-1,\,1]$. The red star represents the SM prediction as in 
Eqs.~(\ref{SpsiphiSMPrediction}) and (\ref{AbSLSMPrediction}), that is between the $2\sigma$ and 
$3\sigma$ levels. The figure shows that only small deviations from the SM value are allowed, corresponding 
to large values of $a_{CP}$: such deviations go in the right direction, but are too small to reach the 
experimental central value.}
\item{In plot (b) of Fig.~\ref{fig:AbSLcorrelation}, we show the MFV scenario in which $a_{CP}=0$, and 
$a_W$ and $a_Z^d$ are non-vanishing. Only few points survive and they all are quite close to the SM value: 
the reason can be understood looking at Fig.~\ref{fig:EpsilonKRatio}(c), where indeed $a_W$ is allowed 
to span only a narrow range $\sim[-0.1,\,0.1]$, corresponding to tiny contributions to $A^b_{sl}$.}
\item{Finally, in plot (c) of Fig.~\ref{fig:AbSLcorrelation}, $a_W$ and $a_{CP}$ are both allowed to take 
values in the $[-1,\,1]$ range. Tiny, but visible, deviations of $S_{\psi\phi}$ from the SM value are now 
allowed, but this is not the case for $A^b_{sl}$. This can be understood from Eqs.~(\ref{CBds}) and 
(\ref{NPphaseB}): when $a_W$ and $a_{CP}$ take the same values but opposite in sign, as it is the case 
here (compare with Fig.~\ref{fig:EpsilonKRatio}(e)), then a cancellation occurs in $C_{B_{d,s}}$ and 
therefore $\delta A^b_{sl}\approx0$; such cancellation, however, does not take place in $\varphi_{B_{d,s}}$ 
and then $\delta S_{\psi\phi}\neq0$.}
\end{enumerate}
Summarizing, in the MFV scenario with a strongly interacting Higgs sector explored in this paper, 
only tiny corrections to the SM predictions for $S_{\psi\phi}$ and $A^b_{sl}$ occur. This outcome 
is still experimentally allowed, mainly due to the large uncertainties which affect present measurements. 
Hopefully, LHCb will (partially) lower such uncertainties in the next future and this will further test 
the parameter space of our scenario.

\section{Conclusions}

Naturalness arguments about the electroweak sector of the SM strongly point to new dynamics around the TeV scale. On the other side, the experimentally allowed gap gets narrowing down for BSM models which naturally propose new light states, such as for instance the simplest supersymmetric  models of the electroweak scale. In this situation, the possibility of a strong-interacting dynamics associated to the electroweak breaking sector regains new interest, be it in the simplest case of a TeV or heavier Higgs particle -an option still allowed by present LHC data within $3\sigma$-  or when the strong dynamics is accompanied of a light Higgs mode.
In both cases, it may well be that the new physics will show up first through its impact on non-standard  fermion couplings and flavour effects. 

Much as it is happening in direct Higgs searches, departures from the SM are neither being found in flavour physics, which is an optimal  tool to indirectly detect new physics. The latter fact led to the fructiferous MFV ansatz. 
In this work, we have identified the leading effective couplings which would signal MFV for the case of a strong-interacting dynamics associated to the Higgs sector. The analysis does not refer to any particular BSM theory: an effective Lagrangian approach has been used, with the strong-Higgs dynamics parametrized as usual through a generic non-linear sigma model. The leading operators appear at mass dimension four, and we have allowed in the analysis custodial-breaking couplings, within their strong phenomenological bounds.

In the linear realization of MFV there were four effective operators involving quarks and the Higgs field 
at leading order (i.e. dimension six). The non-linear realization considered here turns out to have as well 
four operators of that type at its leading order (i.e. dimension four). But only two among the latter are 
``siblings" of two of the former set, in the sense that their low-energy impact on flavour observables 
is the same. The nature of the other two leading couplings differs for both regimes. In particular, one 
outstanding difference of MFV in the strong interacting Higgs scenario, compared with the purely perturbative 
regime, is the natural presence of a new source of CP violation at leading  order of the effective field theory. 

We have analyzed in depth the phenomenological constraints and hypothetical future impact of the 
leading effective couplings. All their contributions to low-energy measurements depend only on three 
parameters: the modification of the Z-fermion couplings for down quarks\footnote{Because of the Yukawa 
dependence dictated by the MFV hypothesis, the  Z-fermion couplings for up quarks $a_Z^u$ can be neglected} 
$a_Z^d$, and of the W-fermion couplings, $a_W$ and $a_{CP}$ for their real and imaginary parts, respectively. 
Because the corrections are tantamount to a non-unitary component in flavour mixing, new CP-odd effects 
follow even when only two fermion families are present, in contrast to the SM case.

In the linear realizations of MFV, either the operator coefficients are assumed to be $~\cO(1)$ and 
the new physics scale needs to be larger than 5 TeV, or the flavour scale is assumed to be $\sim1$ TeV, 
in which case  the fermion-boson couplings must be small. In contrast, in the case of strongly-interacting 
Higgs sector explored here, the scale of the dimension four operators analyzed is necessarily the mass 
of the W and Z gauge bosons: as a consequence,  the experimental constraints on $\Delta F=1$ processes 
force $a_Z^d$ to be small, as for linear MFV, while $a_W$ and $a_{CP}$ can be even large. This is a big 
difference with respect to the linear case, because it is now easier to have correlations and in particular 
cancellations among different contributions:  this happens for instance for $\epsilon_K$ and other 
observables, see Fig.~3.

A plethora of $\Delta F=1$ and $\Delta F=2$ processes have been analyzed in much detail in this work, 
including the existing  tension between the exclusive and the inclusive experimental determinations 
of $|V_{ub}|$, which within the SM translates into the $\vep_K-S_{\psi K_S}$ anomaly and also the 
$BR(B^+\to\tau^+\nu)$ anomaly. In particular, the non-linear realization of MFV is able to remove 
the $\vep_K-S_{\psi K_S}$ anomaly, while it does not soften the SM tension for $BR(B^+\to \tau^+\nu)$.
For the reference values $(|V_{ub}|,\gamma)=(3.5\times 10^{-3},\,66^\circ)$, the  new physics parameter 
space for the ensemble of observables $\vep_K$, $S_{\psi K_S}$, the ratio $BR(B^+\to\tau^+\nu)/\Delta 
M_{B_d}$,  and $\Delta M_{B_d}/\Delta M_{B_s}$ turns out to be  strongly constrained. Only for a small 
part of it all the observables take values inside their own experimental $3\sigma$ errors ranges; when 
the type of cancellations mentioned above occur, $a_W$ and $a_{CP}$ may be even large. Furthermore, 
tiny corrections to the SM predictions for the CP asymmetries $S_{\psi\phi}$ and $A^b_{sl}$ may occur, 
and this is still experimentally allowed. Hopefully LHCb will (partially) lower these experimental 
uncertainties in the near future,  further constraining the parameter space of our scenario or heralding 
a signal of new physics. 
 
We are living exciting times in which new data appear galore from several fronts; flavour physics 
is offering an  increasingly piercing window on new physics and it could even shed the first light 
on a possible dynamic nature of the electroweak symmetry breaking mechanism itself.


\section*{Acknowledgments}  
L.M. thanks Andrzej J. Buras for interesting discussions and Paride Paradisi and Emmanuel Stamou for 
their valuable comments.  L.M. thanks the Aspen Center for Physics for kind hospitality during the 
preparation of this work. R. Alonso acknowledges financial support from the MICINN grant BES-2010-037869. 
The work of J. Y. is supported by the European Programme Unification in the LHC Era (UNILHC), under the 
contract PITN-GA-2009-237920. R.A., M.B.G. and S.R. acknowledge CICYT support through the project 
FPA-2009 09017. R.A., M.B.G. and J.Y. acknowledge CAM support through the project HEPHACOS, P-ESP-00346. 
R. A., M.B.G., S. R. acknowledge partial support from the European Program ``Unification in the LHC era" 
under contract PITN- GA-2009-237920 (UNILHC). M.B.G. is indebted to Gran Wyoming (El Intermedio) and 
his hymn ``Resistir\'e" for inspiration.


\appendix

\section{Useful Formulas for non-linear $d_{\chi}=4$ basis}

In this appendix the derivation of $d_{\chi}=4$ operators is sketched. To make contact with all the 
strong Higgs effective Lagrangian literature the relation between the notation used in this paper and 
the one also used in the literature (see for example \cite{Longhitano:1980tm}) is shown. In this appendix, 
only operators involving fermions and the strong Higgs sector are analyzed. For the complete basis of 
operators, including all the gauge-strong Higgs interactions, one can refer to \cite{Appelquist:1980vg,
Longhitano:1980iz,Longhitano:1980tm,Appelquist:1984nc,Appelquist:1984rr,Cvetic:1988ey,DeRujula:1991se,
Feruglio:1992fp}.

The main quantities needed in the construction are the basic ``covariant'' quantities under the 
SM gauge group:
\beq
\begin{aligned}
\TL &=  \UH\tau_3 \UH^{\dagger}\,,\qquad\qquad
&&\,\,\,\TL \raw L\, \TL L^\dagger\,, \\
\VL_\mu &=  ({\cal D}_\mu \UH) \UH^{\dagger}\,,\qquad\qquad 
&&\VL_\mu \raw L\, \VL_\mu L^\dagger\,, 
\label{VT}
\end{aligned}
\eeq
where for historical reasons $L\in SU(2)_L$ and $R\in U(1)_Y$ are denoted respectively.
It is straightforward to verify that all these quantities are traceless:
\be
\tra\left[ \TL \right] = \tra \left[ \VL_\mu \right] = 0 \,,
\label{VTcomponents}
\ee
and consequently, using the well known decomposition properties of a generic $2\times 2$ matrix, one can 
therefore decompose them as:
\beq
\TL = \frac{1}{2} \tra\left[ \TL \tau^i \right] \tau_i\,,\qquad\qquad
\VL_\mu = \frac{1}{2} \tra\left[ \VL_\mu \tau^i \right] \tau_i
\label{TLVL} \,.
\eeq
All operators, at any order in the chiral expansion, can be expressed in terms of the vectors $T^i$, 
$V^i_\mu$ or of the corresponding traces. For sake of completeness we report here the explicit 
expression of the relevant traces in the unitary gauge:
\beq
\begin{aligned}
\tra\left[ \TL \tau^i \right]_U &= 2 \, \delta_{i3} \qquad\qquad 
&&{\rm for}~i=1,2,3  \\ 
\tra\left[ \VL_\mu \tau^a \right]_U &= i g W^a_\mu  \qquad\qquad 
&&{\rm for}~a=1,2  \\
\tra\left[ \TL \VL_\mu \right]_U &= i \frac{g}{c_W} Z_\mu \,. 
\end{aligned}
\eeq

\boldmath
\subsection*{$d_{\chi}=4$ Contributions}
\unboldmath

Being $\TL$ an hermitian and unitary matrix, it is straightforward to show that all the Higgs ``covariant'' 
quantities with a single Lorentz index, that can be built starting from Eq.~(\ref{TLVL}) are given by: 
\beq
\begin{aligned}
\VL_\mu &= \frac{1}{2} \tra\left[ \VL_\mu \tau^i \right] \tau_i\,, \\
\left(\TL\VL^\mu +\VL^\mu\TL\right) &= \tra\left[ \TL \VL_\mu \right] \mathbb{1}\,,  \\
\left(\TL\VL^\mu - \VL^\mu\TL\right) &= \frac{i}{2}\epsilon^{ijk}\,\tra\left[\TL\tau_i\right]
     \tra\left[\VL^\mu\tau_j\right]\,\tau_k\,, \\
\TL \VL_\mu \TL &= \frac{1}{2} \left( \tra\left[ \TL \VL_\mu \right] \tra\left[ \TL \tau^i \right] - 
              \tra\left[ \VL_\mu \tau^i \right] \right) \tau_i \,.                  
\end{aligned}
\label{combVT}
\eeq
Using the relations in Eqs.~(\ref{VT})-(\ref{combVT}) the operators defined in Eqs.~(\ref{siblings})-
(\ref{nosiblings}) can be written, alternatively, as:
\bea
\mathcal{O}_{1} &=& \frac{1}{2}J^{\mu} \,\tra[\TL \VL_\mu ]\,,  \label{Op1} \\ 
\mathcal{O}_{2} &=& \frac{1}{2}J^{\mu}_i \, \tra[\VL_{\mu}\tau^i]\,,  \label{Op2} \\  
\mathcal{O}_{3} &=& \frac{1}{2}J^{\mu}_i \, \left(\tra[\TL \VL_{\mu}]\,\tra[\TL\tau^i]-
                    \tra[\VL_\mu\tau^i]\right)\,, \label{Op3} \\
\mathcal{O}_{4} &=& \frac{i}{4}\eps^{ijk}\,\tra[\TL\tau_i]\,\tra[\VL_{\mu}\tau_j] J^{\mu}_k \,,\label{Op4}
\label{CPodd}
\eea
with $J^\mu$ and $J^\mu_i$ the $SU(2)_L$ singlet and triplet currents, respectively:
\beq
J^{\mu}=i\bar{Q}_L \lambda_{F} \gamma^{\mu} Q_L\,,\qquad \qquad 
J^{\mu}_{i} = i \bar{Q}_L\lambda_{F}\gamma^{\mu} \tau_i Q_L \, .
\eeq

One of the most relevant differences of the strong interacting Higgs scenario, when compared with 
the linear case is the presence of a new source of CP violation at chiral dimension $d_\chi=4$. 
To realize explicitly the CP character of non-linear operators in the strong Higgs case one has 
to remember the CP transformation properties \cite{Longhitano:1980tm} of $\TL$ and $\VL_\mu$:
\bea
\TL(t,x) \quad &\stackrel{\mathcal{CP}}{\longrightarrow} &\quad -\tau_2 \,\TL(t,-x) \,\tau_2\,,\label{TCP} \\
\VL_\mu(t,x)\quad &\stackrel{\mathcal{CP}}{\longrightarrow} &\quad\phantom{-}\tau_2\,\VL^\mu(t,-x) \,\tau_2\,.\label{VCP}
\eea
By means of Eqs.(\ref{TCP}) and (\ref{VCP}) it is straightforward to recover the transformation properties 
of the traces:
\bea
\tra[\TL \tau_i] \quad &\stackrel{\mathcal{CP}}{\longrightarrow} & \quad \phantom{-}\tra[\TL \tau_i^*]\,, \\
\tra[\VL_\mu \tau_i] \quad &\stackrel{\mathcal{CP}}{\longrightarrow} & \quad - \tra[\VL^\mu \tau_i^*]\,,  \\
\tra[\TL \VL_{\mu}] \quad &\stackrel{\mathcal{CP}}{\longrightarrow} & \quad -\tra[\TL\VL^\mu] \, .
\eea
Using in addition the transformation properties of the singlet and triplet $SU(2)_L$ fermionic currents:
\bea
\overline{Q}_L\gamma^{\mu} Q_L \quad &\stackrel{\mathcal{CP}}{\longrightarrow} & 
           \quad - \bar{Q}_L\gamma_{\mu} Q_L\,, \\
\overline{Q}_L\gamma^{\mu} \tau_i Q_L \quad &\stackrel{\mathcal{CP}}{\longrightarrow} & 
           \quad - \bar{Q}_L\gamma_{\mu} \tau_i^*Q_L\,,
\eea
one can easily verify that $\mathcal{O}_{1,2,3}$ are CP-even, while $\mathcal{O}_{4}$ is CP-odd.

\subsection*{Relation With the Linear Representation}

Finally we want to connect the operators in the non-linear basis with those defined in the linear 
realization. One can define the following four flavour operators involving fermions and Higgs fields:
\bea
\cO_{H1} &=& i\left(\overline{Q}_L\lambda_{FC}\gamma^\mu Q_L\right)
          \left(\Phi^\dagger (D_\mu\Phi)-\left(D_\mu\Phi\right)^\dagger \Phi\right)\,, \label{RelSt-Lin1} \\
\cO_{H2} &=& i\left(\overline{Q}_L\lambda_{FC}\gamma^\mu\tau^i Q_L\right)\left(\Phi^\dagger
        \tau_i (D_\mu\Phi)-\left(D_\mu \Phi\right)^\dagger \tau_i\Phi\right)\,,\label{RelSt-Lin2} \\
\cO_{H3} &=& i\left(\overline{Q}_L\lambda_{FC}\gamma^\mu\tau^i Q_L\right)\left(\Phi^\dagger\tau_i 
        \Phi\right) \left(\Phi^\dagger (D_\mu\Phi)-\left(D_\mu\Phi\right)^\dagger \Phi\right)\,, \\
\cO_{H4} &=& i\eps^{ijk}\left(\overline{Q}_L\lambda_{FC}\gamma^\mu\tau_i Q_L\right)\left(\Phi^\dagger\tau_j 
        \Phi\right)\left(\Phi^\dagger\tau_k (D_\mu\Phi)-\left(D_\mu\Phi\right)^\dagger\tau_k\Phi\right)\,.
\label{RelSt-Lin4}
\eea
The first two operators \cite{D'Ambrosio:2002ex} appear in the linear expansion at dimension $d=6$, 
while the last two appear only at dimension $d=8$. To match the operators in the-linear and non 
linear expansion one has to express the traces appearing in Eqs.~(\ref{Op1})-(\ref{Op4}) with 
the corresponding (linear) Higgs contribution. In particular one finds: 
\bea 
\tra[\TL \VL_{\mu}] \quad & \raw & \quad - \frac{2}{v^2} 
     \left(\Phi^{\dagger}(D_{\mu}\Phi)-\left(D_{\mu}\Phi\right)^{\dagger}\Phi\right) \label{NLtoL1} \\
\tra[\VL_{\mu}\tau^i] \quad &\raw & \quad \phantom{-}\frac{2}{v^2}\left(\Phi^{\dagger}\tau^i(D_{\mu}\Phi)-
     \left(D_{\mu}\Phi\right)^{\dagger}\tau^i \Phi\right) \label{NLtoL2} \\ 
\tra[\TL \tau^i] \quad &\raw &\quad -\frac{4}{v^2} (\Phi^{\dagger}\tau^i \Phi)  \, . \label{NLtoL3}
\eea
Here the correspondence between $\langle M^{\dagger} M \rangle =v^2$ and $\langle \Phi^{\dagger} 
\Phi \rangle = v^2/2$ has to be kept in mind. Inserting Eqs.~(\ref{NLtoL1})-(\ref{NLtoL3}) in 
Eqs.~(\ref{Op1})-(\ref{Op4}) one can easily recover the following correspondence between operators 
in the linear and non linear realization:
\begin{flalign}
&\mathcal{O}_{1}\leftrightarrow -\frac{1}{v^2}\mathcal{O}_{H1} \, , \\
&\mathcal{O}_{2}\leftrightarrow  \frac{1}{v^2}\mathcal{O}_{H2} \, , \\
&\mathcal{O}_{3}\leftrightarrow  \frac{4}{v^4}\mathcal{O}_{H3} - \frac{1}{v^2}\mathcal{O}_{H2} \, , \\
&\mathcal{O}_{4}\leftrightarrow -\frac{2}{v^4}\mathcal{O}_{H4} \, .
\end{flalign}
In particular notice that $d_{\chi}=4$ operators $\mathcal{O}_{1,2}$ correspond to the $d=6$ 
linear operators $\cO_{H1}$ and $\cO_{H2}$, whereas $\cO_{3,4}$ to operators of dimension up 
to $d=8$ in the linear expansion.

%
%

\section{Formulae for the Phenomenological Analysis}
\label{App:PhemSection}

In this appendix we provide details on the results presented in sects.~\ref{sec:DeltaF2SemileptonicAsym} and \ref{sec:PhenoAnalysis}.

\mathversion{bold}
\subsection{$\Delta F=2$ Wilson Coefficients}
\mathversion{normal}

The Wilson coefficients of the $\Delta F=2$ observables in presence of NP can be written separating the contributions from the box diagrams and the tree-level $Z$ diagrams, so that
\beq
C^{(M)}=\Delta_{Box}^{(M)}\,C+\Delta_Z^{(M)}\,C\,,
\eeq
where $M=K,\,B_d,\,B_s$. Taking into account only the corrections to the $W$-quark vertices, we find the following contributions to the Wilson coefficient relevant for $M^0-\bar M^0$ system at the matching scale $\mu_t\approx m_t(m_t)$ ($m_t(m_t)$ is the top quark mass $m_t$ computed at the scale $m_t$ in the $\ov{\mathrm{MS}}$ scheme):
\beq
\Delta_{Box}^{(M)}\,C(\mu_t)=\sum_{i,j=u,c,t}\tilde\la_i\,\tilde\la_j\,F_{ij}\,,
\label{DeltaCbox}
\eeq
where for the $K$ and $B_q$ systems we have respectively
\beq
\tilde\la_i=\tilde V^*_{is}\,\tilde V_{id}\,,\qquad\qquad
\tilde\la_i=\tilde V^*_{ib}\,\tilde V_{iq}\,,
\label{tildelambdaCKM}
\eeq
with $\tilde V$ the modified CKM matrix. The $F_{ij}$ are the usual box functions with the exchange of $W$ and up-type quarks defined by
\beq
\begin{aligned}
&F_{ij}\equiv F(x_i,\,x_j) =\dfrac{1}{4}\,\Bigg[(4+x_i\,x_j)\,I_2(x_i,\,x_j)-8\,x_i\,x_j\,I_1(x_i,\,x_j)\Bigg]\\
&I_1(x_i,\,x_j) = \dfrac{1}{(1-x_i)(1-x_j)} + \Bigg[\dfrac{x_i\, \ln(x_i)}{(1-x_i)^2 (x_i-x_j)} + (i\leftrightarrow j)\Bigg]\\
&I_2(x_i,\,x_j) = \dfrac{1}{(1-x_i)(1-x_j)} + \Bigg[\dfrac{x_i^2\, \ln(x_i)}{(1-x_i)^2 (x_i-x_j)} + (i\leftrightarrow j)\Bigg]\,,
\end{aligned}
\eeq
with $x_i=(m_i/M_{W})^2$ (with $m_i$ should be understood as $m_i(m_i)$).

In the SM limit, i.e. switching off the modifications of the $W$-quark couplings, $\tilde V\rightarrow V$ and therefore $\tilde\la_i\rightarrow \la$ and it is possible to rewrite the previous expression in eq.~(\ref{DeltaCbox}) in terms of the usual Inami-Lim functions $S_0(x_t)$, $S_0(x_c)$ and $S_0(x_c,\,x_t)$, using the unitarity relations of the CKM matrix:
 \beq
\begin{aligned}
&S_0(x_t)\equiv\dfrac{4\,x_t-11\,x_t^2+x_t^3}{4(1-x_t)^2}-\dfrac{3\,x_t^3\,\log x_t}{2(1-x_t)^3}\\
&S_0(x_c)\equiv x_c\\
&S_0(x_c,\,x_t)\equiv x_c\left[\log \dfrac{x_t}{x_c}-\dfrac{3\,x_t}{4(1-x_t)}-\dfrac{3\,x_t^2\,\log x_t}{4(1-x_t)^2}\right]\,.
\end{aligned}
\eeq

When the NP contributions are taken into account, it is useful to analyse the modification of the CKM factors $\lambda_i$ entering the meson oscillation Wilson coefficients: for the $K$ system we have
\bea
\tilde \lambda_i &\equiv & \tilde{V}_{id_2}^*\,\tilde{V}_{id_1}=\lambda_i\,\left[1+i\,a_{CP}\,\Delta^d_{12}
       +a_W\,(\Sigma^d_{1i}+\Sigma^d_{2i})+(a^2_{W}+a_{CP}^2)\, \Sigma^d_{1i} \Sigma^d_{2i}\right] \, , \\
\tilde \lambda'_i &\equiv &\tilde{V}_{u_2 i}^*\,\tilde{V}_{u_1 i}=\lambda'_i\,\left[1+i\,a_{CP}\,\Delta^u_{12} 
      +a_W\,(\Sigma^u_{1i}+\Sigma^u_{2i})+(a^2_{W}+a_{CP}^2)\, \Sigma^u_{1i} \Sigma^u_{2i}\right] \, ,
\eea
with $\lambda_i=V_{i d_2}^*\,V_{i d_1}$, $\lambda'_i=V_{u_2 i}^*\,V_{u_1 i}$ and
\be
\Delta^x_{12} = y_{x_1}^2-y_{x_2}^2 \,,\qquad\qquad 
\Sigma^x_{1i} = y_{x_1}^2+y_i^2 \,,\qquad\qquad 
\Sigma^x_{2i} = y_{x_2}^2+y_i^2 \, .
\ee
So respectively for the $B_s$, $B_d$ and $K$ system one has to replace accordingly the quark labels.
If now we consider the products $\tilde \lambda_i\,\tilde\lambda_j$, we can parametrise the deviations from the SM expression as follows: 
\beq
\tilde \lambda_i\,\tilde\lambda_j=\lambda_i\,\lambda_j\,(1+\delta \lambda_{ij})
\label{tildelambdaCKMapp}
\eeq
where for the $K$ system
\beq
\begin{split}
\delta \lambda_{ij}=&2\,i\,a_{CP}\,\Delta^x_{12}+a_{W}\,A^x_{ij}
+a_{CP}^2\,B^x_{ij}+a_{W}^2\,C^x_{ij}+i\,a_{CP}\,a_W\,\Delta^x_{12}\,B^x_{ij}+\\
&+(a_{CP}^2+a_W^2)(i\,a_{CP}\,D^x_{ij}+a_{W}\,E^x_{ij})
+(a_{CP}^2+a_W^2)^2\,L^x_{ij}\,,
\end{split}
\eeq
with
\beq
\begin{aligned}
&A^x_{ij}=\Sigma^x_{1i}+\Sigma^x_{1j}+\Sigma^x_{2i}+\Sigma^x_{2j}\,,\\
&B^x_{ij}=\Sigma^x_{1i}\,\Sigma^x_{2i}+\Sigma^x_{1j}\,\Sigma^x_{2j}-\left(\Delta^x_{12}\right)^2\,,\\
&C^x_{ij}=\Sigma^x_{1i}\,\Sigma^x_{1j}+\Sigma^x_{2i}\,\Sigma^x_{2j}+\Sigma^x_{1i}\,(\Sigma^x_{2i}+\Sigma^x_{2j})+\Sigma^x_{1j}\,(\Sigma^x_{2i}+\Sigma^x_{2j})\,,\\
&D^x_{ij}=\Delta^x_{12}\,(\Sigma^x_{1i}\,\Sigma^x_{2i}+\Sigma^x_{1j}\,\Sigma^x_{2j})\,,\\
&E^x_{ij}=(\Sigma^x_{1i}+\Sigma^x_{2i})\,\Sigma^x_{1j}\,\Sigma^x_{2j}+(\Sigma^x_{1j}+\Sigma^x_{2j})\,\Sigma^x_{1i}\,\Sigma^x_{2i}\,,\\
&L^x_{ij}=\Sigma^x_{1i}\,\Sigma^x_{1j}\,\Sigma^x_{2i}\,\Sigma^x_{2j}\,.
\end{aligned}
\eeq
Notice that the previous expressions hold for both for the $K$ and the for $B_q$ systems: only $\Delta^x_{12}$, $\Sigma^x_{1i}$ and $\Sigma^x_{2i}$ distinguish the different systems. With such notation, we can write the expression in eq.~(\ref{DeltaCbox}) as follows:
\beq
\begin{aligned}
&\begin{split}
\Delta_{Box}^{(K)}\,C(\mu_t)=\la_t^2\,S'_0(x_t)+\la_c^2\,S'_0(x_c)+2\,\la_t\,\la_c\,S'_0(x_c,\,x_t)\,,
\end{split}\\
&\Delta_{Box}^{(B_q)}\,C(\mu_t)=\la_t^2\,S'_0(x_t)\,,
\end{aligned}
\eeq
where
\beq
\begin{aligned}
&S'_0(x_t)\equiv (1+\delta\la_{tt})\,F_{tt}+(1+\delta\la_{uu})\,F_{uu}-2\, (1+\delta\la_{ut})\,F_{ut}\,,\\
&S'_0(x_c)\equiv (1+\delta\la_{cc})\,F_{cc}+(1+\delta\la_{uu})\,F_{uu}-2\, (1+\delta\la_{uc})\,F_{uc}\,,\\
&S'_0(x_c,\,x_t)\equiv (1+\delta\la_{ct})\,F_{ct}+ (1+\delta\la_{uu})\,F_{uu}- (1+\delta\la_{uc})\,F_{uc}- (1+\delta\la_{ut})\,F_{ut}\,.
\end{aligned}
\eeq
When all the $\delta\la_{ij}$ are put to zero, these relations coincide with the SM ones.

Moving to the tree-level FCNC $Z$ diagrams and integrating the $Z$ boson at $\mu_t$~\footnote{Integrating out the $Z$ boson at $\mu_t$ or at $M_Z$ introduces a subleading error in our computation, that can be safely neglected.}, we find the following contributions to the Wilson coefficients:
\beq
\Delta C_Z^{(K)}(\mu_t)=\dfrac{4\,\pi^2}{G_F^2\,M_W^2}\dfrac{1}{2\,M_Z^2}\,(C^{d,s})^2\,,\qquad\qquad
\Delta C_Z^{(B_q)}(\mu_t)=\dfrac{4\,\pi^2}{G_F^2\,M_W^2}\dfrac{1}{2\,M_Z^2}\,(C^{q,b})^2\,,
\label{ZWilsonCoefficients}
\eeq 
where the coefficients $C^{d,s}$ and $C^{q,b}$ are the FCNC couplings of the $Z$ boson to the down-type quarks,
\beq
C^{d,s}=\dfrac{g}{2\cos\theta_W}\,a_Z^d\,(\la_{FC})^*_{12}\,,\quad
C^{d,b}=\dfrac{g}{2\cos\theta_W}\,a_Z^d\,(\la_{FC})^*_{13}\,,\quad
C^{s,b}=\dfrac{g}{2\cos\theta_W}\,a_Z^d\,(\la_{FC})^*_{23}\,.
\eeq

The Wilson coefficients given above are evaluated at the $\mu_t$ scale and therefore the complete analysis requires the inclusion of the renormalisation group (RG) QCD evolution down to low energy scales, at which the hadronic matrix elements are evaluated by lattice methods.

In our model we can apply the same RG QCD analysis as in the SM context: indeed, both the effective operator, that arises from the modified box diagrams and the tree-level $Z$ diagrams by integrating out the heavier degrees of freedom, and the matching scale are the same as in the SM. In particular no new effective operators with different chiral structure from that one in eq.~(\ref{EffectiveOperatorMesonOscillations}) and no higher scales than $\mu_t$ are present. All the NP effects are encoded into the Wilson coefficients.

By the use of the Wilson coefficients reported in this appendix and having in mind the previous discussion on the QCD evolution, we find the following full expressions for the mixing amplitudes:
\beq
\begin{aligned}
&M_{12}^K=R_K\Big[\eta_2\,\la_t^2\,S'_0(x_t)+\eta_1\,\la_c^2\,S'_0(x_c)+
2\,\eta_3\,\la_t\,\la_c\,S'_0(x_c,\,x_t)
+\eta_2\, \Delta C_Z^{(K)}(\mu_t)
\Big]^*\,,\\
&M_{12}^q=R_{B_q}\Big[
\la_t^2\,S'_0(x_t)+\, \Delta C_Z^{(B_q)}(\mu_t)
\Big]^*\,.
\end{aligned}
\label{M12NPfull}
\eeq

Analogously, in the presence of NP also $\Gamma_{12}^q$ is modified. Following Ref.~\cite{Beneke:2003az}, we can write $\Gamma^q_{12}$ in our model as 
\beq
\Gamma_{12}^q=-\left[\tilde\lambda_c^2\,\Gamma_{12}^{cc}+
2\,\tilde\lambda_c\,\tilde\lambda_u\Gamma_{12}^{cc}+
\tilde\lambda^2_u\Gamma_{12}^{uu}\right]\,,
\eeq
where $\tilde\lambda_i$ are the CKM factors defined in eq.~(\ref{tildelambdaCKM}) and $\Gamma_{12}^{ij}$ can be found in Ref.~\cite{Beneke:2003az}. By using the same notation as in eq.~(\ref{tildelambdaCKMapp}) and using the unitarity of $\lambda_i$ to eliminate $\lambda_c$, we get
\beq
\Gamma_{12}^q=-\left[\lambda_t^2\,\Gamma_{12}^{\prime cc}+
2\,\lambda_t\,\lambda_u\left(\Gamma_{12}^{\prime cc}-\Gamma_{12}^{\prime uc}\right)+
\lambda^2_u\left(\Gamma_{12}^{\prime cc}-2\Gamma_{12}^{\prime uc}+\Gamma_{12}^{\prime uu}\right)
\right]\,,
\eeq
where
\beq
\Gamma_{12}^{\prime ij}\equiv \left(1+ \delta\lambda_{ij}\right)\,\Gamma_{12}^{ij}\,.
\eeq
Notice that in the limit in which $\delta\lambda_{ij}\to0$, the SM expression is recovered.

\subsection{Approximate Analytical Expressions}

Once we consider only the first terms in the expansion in $a_W$ and $a_{CP}$, the relevant parameters $\delta\lambda_{ij}$ are simplified as follows: for the $K$ system they are
\beq
\begin{aligned}
&\delta\lambda_{uu}=2\,(a_W-i\,a_{CP})\,y_s^2\,,\\
&\delta\lambda_{cc}=4\,a_W\,y_c^2-2\,i\,a_{CP}\,y_s^2\,,\\
&\delta\lambda_{tt}=4\,a_W\,y_t^2-2\,i\,a_{CP}\,y_s^2\,,\\
&\delta\lambda_{uc}=2\,a_W\,y_c^2-2\,i\,a_{CP}\,y_s^2\,,\\
&\delta\lambda_{ut}=\delta\lambda_{ct}=2\,a_W\,y_t^2-2\,i\,a_{CP}\,y_s^2\,,\\
\end{aligned}
\label{deltalambdaK}
\eeq
while for the $B_q$ systems they are
\beq
\begin{aligned}
&\delta\lambda_{uu}=\delta\lambda_{cc}=\delta\lambda_{uc}=2\,(a_W-i\,a_{CP})\,y_b^2\,,\\
&\delta\lambda_{tt}=2\,a_W\,(2\,y_t^2+y_b^2)-2\,i\,a_{CP}\,y_b^2(1+2\,a_W\,y_t^2)\,,\\
&\delta\lambda_{ut}=\delta\lambda_{ct}=2\,a_W\,(y_t^2+y_b^2)-2\,i\,a_{CP}\,y_b^2(1+a_W\,y_t^2)\,,\\
\end{aligned}
\label{deltalambdaB}
\eeq
where the subsequent terms in the expansion are of order $\cO(a_W^2,\,a_{CP}^2,\,a_W\,a_{CP})$.
Similarly, we can report the approximated expressions for the $C^{d,s}$, $C^{d,b}$ and $C^{s,b}$ couplings, that enter the tree level $Z$ contributions:
\beq
\begin{aligned}
C^{d,s}=\dfrac{g}{2\cos\theta_W}\,a_Z^d\,y_t^2\,V_{ts}^*\,V_{td}\,,\\
C^{d,b}=\dfrac{g}{2\cos\theta_W}\,a_Z^d\,y_t^2\,V_{tb}^*\,V_{td}\,,\\
C^{s,b}=\dfrac{g}{2\cos\theta_W}\,a_Z^d\,y_t^2\,V_{tb}^*\,V_{ts}\,.
\end{aligned}
\eeq\\

Finally we report the explicit expressions for $G(x_i)$ and $H(x_i,\,x_j)$ appearing in the expressions in sect.~\ref{sec:DeltaF2SemileptonicAsym}:
\beq
\begin{aligned}
&G(x_t)=2(F_{tt}-\,F_{ut})=\dfrac{4\,x_t+2\,x_t \log x_t}{1-x_t}-\dfrac{7\,x_t^2-x_t^3}{2\,(1-x_t)^2}+\dfrac{2\,x_t^2-5\,x_t^3}{(1-x_t)^3}\log x_t\,,\\
&G(x_c)=2(F_{cc}-\,F_{uc})=2\,x_c\,(2+\log x_c)\,\\
&H(x_t,x_c)=F_{tc}-F_{ut}=x_c\left(\dfrac{4-11\,x_t+7\,x_t^2}{4\,(1-x_t)^2}+\dfrac{4-8\,x_t+x_t^2}{4\,(1-x_t)^2}\log x_t\right)\,,\\
&H(x_c,x_t)=F_{ct}-F_{uc}=-x_c\,\log x_c+x_t\dfrac{4-3\,x_c}{4\,(1-x_t)}+\dfrac{4\,x_t+x_c(4-8\,x_t+x_t^2)}{4\,(1-x_t)^2}\log x_t\,.
\end{aligned}
\eeq


\providecommand{\href}[2]{#2}\begingroup\raggedright\endgroup

\end{document}